\documentclass[aps,amsmath,amssymb,twocolumn,showpacs,superscriptaddress,pra]{revtex4-1}

\usepackage{mathrsfs}
\usepackage{leftidx}
\usepackage{graphicx, subfigure}  
\usepackage{pgfplotstable, pgfplots} 
\usepackage{dcolumn}   
\usepackage{bm}        
\usepackage{braket}
\usepackage{comment} 
\usepackage{amssymb}   
\usepackage{enumerate}
\usepackage{amsmath}
\usepackage{float}
\usepackage[colorlinks,
            linkcolor=red,
            anchorcolor=blue,
			urlcolor=blue,
            citecolor=green]{hyperref}

\begin{document}


\title{Symmetry-breaking dynamics of the finite-size Lipkin-Meshkov-Glick model near ground state}

\author{Yi Huang}
\affiliation{Center for Quantum Information, Institute for Interdisciplinary Information Sciences, Tsinghua University, Beijing 100084, China}
\affiliation{Department of Applied Physics, Xi'an Jiaotong University, Xi'an, Shaanxi 710049, China}

\author{Tongcang Li}
\affiliation{Department  of  Physics  and  Astronomy, Purdue  University,  West  Lafayette,  IN  47907,  USA}
\affiliation{School  of  Electrical  and  Computer  Engineering, Purdue  University,  West  Lafayette,  IN  47907,  USA}
\affiliation{Purdue  Quantum  Center,  Purdue  University,  West  Lafayette,  IN  47907,  USA}
\affiliation{Birck Nanotechnology Center,  Purdue University,  West Lafayette,  IN 47907,  USA}

\author{Zhang-qi Yin}\email{yinzhangqi@tsinghua.edu.cn}
\affiliation{Center for Quantum Information, Institute for Interdisciplinary Information Sciences, Tsinghua University, Beijing 100084, China}

\date{\today}

\begin{abstract}

We study the dynamics of the Lipkin-Meshkov-Glick (LMG) model with finite number of spins. In the thermodynamic limit, the ground state of the LMG model with isotropic Hamiltonian
in broken phase breaks to a mean-field ground state with a certain direction. However, when the spins number
$N$ is finite, the exact ground state is always unique and is not given by a classical mean-field ground state. Here we prove that for $N$ is large but finite, through a tiny external perturbation, a localized state which is close to a mean-field ground state can be prepared, which mimics a spontaneous symmetry breaking (SSB). Besides, we find the localized in-plane spin polarization oscillates with two different frequencies $\sim O(1/N)$, and the lifetime of the localized state is long enough to exhibit this oscillation. We numerically test the analytical results and find that they agree with each other very well. Finally, we link the phenomena to quantum time crystals and quasicrystals.

\end{abstract}

\pacs{Valid PACS appear here}
\maketitle


\section{\label{sec:level1}Introduction}
The spontaneous symmetry breaking (SSB) has been widely studied in physics. It is relating to a wide class of phenomena, such as magnetization of ferromagnetic material
(breaking rotational symmetry), crystallization (breaking spherical translational symmetry), and Anderson-Higgs mechanism \cite{Anderson,Higgs} (breaking gauge symmetry).
In 2012, Wilczek proposed an idea of time crystal \cite{Wilczek2012,Shapere2012}, which spontaneously breaks time translational symmetry. Later, Li and
et al. proposed a scheme to realized quantum space-time crystals and quasicrystals in the trapped ions system \cite{Li2012}.
However, it was proven that the original idea of quantum time crystal in the equilibrium state put forward by Wilczek is not possible \cite{Bruno2013a,Bruno2013b,Wilczek2013a,arxiv1212.6959v2}.
In 2015, a no-go theorem was proved to exclude
the existence of quantum time crystals near the ground states for a wide class of Hamiltonian with short range coupling terms \cite{Watanabe2015}.

The idea of time crystals stimulated further studies on the spontaneous temporal symmetry breaking in dissipative systems \cite{Wilczek2013}, or in the systems far from the equilibrium states,
for example, Floquet or discrete time crystals \cite{Sacha2015,PhysRevLett.116.250401,PhysRevLett.117.090402,Yao2017}. In this way, the no-go theorem for quantum time crystals
can be bypassed. It was found that the Floquet time crystal is a type of prethermalization phenomena \cite{PhysRevX.7.011026}.
The length of prethermal regime in time is nearly exponentially long, and the time translational symmetry can be spontaneously broken.
Soon after the theory of discrete time crystal being published, two experiments in trapped ions and nitrogen-vacancy centers were performed
to verified the existence of discrete time crystals \cite{Zhang2017,Choi2017}.

The success in experiments attracted many people to study
discrete quantum time crystals in various many-body systems under Floquet driving \cite{PhysRevB.95.195135,PhysRevLett.119.010602,2017PhRvB..95u4307R,2017arXiv170801472G}.
The possibility of time quasicrystals was also investigated in some classical systems \cite{timequasicrystal,timequasicrystal1}.
The time crystal behavior may also appear in an exited eigenstate of Wilczek's model \cite{Syrwid2017}.
The experimental success also stimulated people to rethink how to realize time crystals near the quantum ground state. For example, combining the ideas of
both classical and quantum time crystals, Shapere and Wilczek proposed a model which realizes the classical time crystal Lagrangians and emergent Sisyphus dynamics around its quantum ground state \cite{Shapere2017}.

In this paper we discuss the spontaneous dynamics of the Lipkin-Meshkov-Glick (LMG) model near its ground state \cite{LIPKIN1965188}.
The LMG model was first proposed to describe the phase transitions in atomic nuclei \cite{LIPKIN1965188}.
The quantum phase transition, the SSB, and the finite size effects in the LMG model have been studied for many years \cite{0305-4470-29-14-023,Dusuel2005,PhysRevE.78.021106,PhysRevLett.93.237204,Quan2007}.
Recently, it is found that the LMG model are relevant to many quantum systems, such as cavity QED \cite{PhysRevLett.100.040403}, optically
trapped nanodiamond with Nitrogen-vacancy centers \cite{MaYue2017}, Bose-Einstein condensation \cite{PhysRevLett.105.204101}, et al.
In the LMG model, spin squeezing \cite{MA201189,PhysRevLett.118.083604}, quantum entanglement \cite{PhysRevLett.101.025701,2013PhLA..377.1053Z}, quantum thermodynamic cycle \cite{PhysRevE.96.022143}, et al. were studied. However, to our best knowledge, there is very little study on the dynamics of the finite-size LMG model.
We find that there is deep connection between the Hamiltonians of the quantum
space-time crystal \cite{Li2012,Wilczek2012} and the LMG model. Thus we expect the behavior near the ground state of the LMG model would be similar to the behavior of quantum time crystals, which is equivalently to say that the expectation value of some physical observables can exhibit a nontrivial dynamics instead of staying still.
\\

This article is structured as follows. Section \uppercase\expandafter{\romannumeral2} discusses the link between the LMG model and space-time crystals based on trapped ions, and the eigenenergy of the LMG Hamiltonian \cite{LIPKIN1965188}. SSB and the dynamics under mean-field approximation of the LMG model are discussed in section \uppercase\expandafter{\romannumeral3}.
For the LMG model with finite but large number of spins, mean-field approximated results show that for localized states near ground state there are oscillations of the in-plane polarizations $S_{x, y}$ with small but nonzero angular frequencies $\sim O(1 / N)$. We construct a trial wave function with lifetime $\sim O(N^3)$ and localization $\sim O(\sqrt{N})$, which can demonstrate the frequency $\sim O(1/N)$ oscillation behavior. In the thermodynamic limit, SSB would localize the ground state but cease the oscillation, which is similar to Wilczek's model.
In section \uppercase\expandafter{\romannumeral4}, the analytical result of the dynamics of the LMG model with finite spins is demonstrated.
We find that there are two intrinsic frequencies for the oscillation of localized states near the ground state.
To understand the spontaneous property of this symmetry-breaking oscillation, we investigate the time correlation function of the finite-size LMG model in section \uppercase\expandafter{\romannumeral5}. The analytical result shows two frequency components in the correlation function match the two intrinsic frequencies just computed in section \uppercase\expandafter{\romannumeral4}. We numerically test the theoretically predicted oscillating behavior in section \uppercase\expandafter{\romannumeral6} by introducing a tiny perturbation. It is found that the numerical curve agrees with the theory very well.
We also discuss the connection between the oscillation in the LMG model and quantum time crystals or quasicrystals. In the last section, we compare our results to time crystal behavior found in the excited eigenstates of Wilczek's model described by Syrwid and et al. \cite{Syrwid2017}, briefly discuss the experimental feasibility, and give a conclusion.

\section{The Lipkin-Meshkov-Glick Model}\label{Hamiltonian}
The original proposals for both time and space-time crystals considered charged particles trapping along a ring trap, under a external magnetic flux $\alpha$
\cite{Wilczek2012,Li2012}. The Hamiltonian is in the form of $H= \frac{1}{2} \sum_i^N (p_i-\alpha)^2 + \text{particles interaction terms}$ \cite{Sacha2017}.
As the ring trap is finite, the value of angular momentum is discrete. As long as $\alpha$ is neither integral nor half integral, the ground state momentum is
not zero. Once the wavefunction of the system is localized, it may start to rotate. However, the spontaneous localization in this system is not well understood.

On the other hand, we have a highly developed theory for SSB in the spin-based models. If the angular momentum operator $p_i$ is replaced with the spin operator $S_z$ with dimension $N+1$, and neglect the particles interaction terms, we find that the original Hamiltonian becomes similar to the Hamiltonian in the isotropic LMG model. Therefore, we anticipate that for the LMG model with large spin number $N$, its ground state may be localized and the in-plane polarization will oscillate spontaneously. In other words, the quantum time crystal may appear in this model.

The Hamiltonian of a general LMG model is
\begin{equation}
	H = \frac{\lambda}{N} (S_x^2 + \gamma S_y^2) - h S_z, \label{h0}
\end{equation}
where $S_{\alpha} = \sum_i \sigma_{\alpha}^i / 2$ is the total spin operator summing over $N$ spins, where $\sigma_{\alpha}$ is the Pauli matrix.
In this paper, we adopt natural unit with $\hbar=1$.
The reason why $N$ is in the denominator is to ensure a finite energy per spin in the thermodynamic limit. The parameter space we are interested in is $\lambda < 0$ standing for a ferromagnetic interaction, $\gamma \in (0, 1]$ describing the anisotropic in-plane coupling, and the magnetic field along $z$ direction with $h \geq 0$.  Also, we confines the following discussion with $\lambda = -1$ for simplicity. In this case, it is well known that a quantum phase transition happens at $h = 1$. We adopt the convention introduced by Dusuel and Vidal \cite{Dusuel2005} to distinguish two phases in the LMG model: symmetric phase featured by $h \geq 1$ and broken phase characterized by $0 \leq h < 1$.

The magnitude of the total spin is conserved
\begin{equation}
	[H, \vec S^2] = 0.
\end{equation}
$S(S + 1)$ is the eigenvalue of $\vec S^2$, and we only concern $S = N / 2$ sector where the lowest-energy states locate. In addition, $H$ is invariant under the following transformation $\sigma_x \rightarrow -\sigma_x$, $\sigma_y \rightarrow -\sigma_y$, and $\sigma_z \rightarrow \sigma_z$, which indicates a $\mathbb {Z} _{2}$ symmetry (spin-flip symmetry). In other words,  $e^{i\pi(S_z - S)}$ or $\prod_i \sigma_z^i$, is also conserved
\begin{equation}
	\left[H, e^{i\pi(S_z - S)}\right] = \left[H, \prod_i \sigma_z^i\right] =  0.
\end{equation}
This $\mathbb {Z} _{2}$ symmetry leads to several consequences such that any eigenstate satisfies
\begin{equation}
	\langle S_x \rangle = \langle S_y \rangle = 0.
\end{equation}

Furthermore, we focus our interest in the isotropic Hamiltonian where $\gamma = 1$, and we expect the rotational symmetry in the $x-y$ plane will make this model exhibit some interesting properties. The Hamiltonian reads
\begin{equation}
	H = - \frac{1}{N} (\vec S^2 - S_z^2) - h S_z.
\end{equation}
$H$ thus commutes both with $\vec S^2$ and $S_z$ so that $H$ is diagonal in the standard eigenstate basis $\{ |S,M \rangle\}$ of $\vec S^2$ and $S_z$. The eigenenergies are
\begin{equation}
	E(S,M) = - \frac{1}{N} [S(S + 1) - M^2] - h M.
\end{equation}

For broken phase, the ground state satisfies
\begin{equation}
	M_0 =
	\left \{
		\begin{array}{ll}
				Int \left[ \frac{Nh}{2} + \frac{1}{2} \right] - \frac{1}{2}, & \text{when N is odd,} \\
				& \\
				Int \left[ \frac{Nh}{2} \right], & \text{when N is even,}
		  \end{array}
	\right.
		  \label{groundm}
\end{equation}
where $Int[ x ]$ gives the integer part of $x$. While for symmetric phase
\begin{equation}
	M_0 = N / 2.
\end{equation}

Since there is very small amplitude of the in-plane polarization in symmetry phase, the next section we will focus on the dynamics in broken phase (ferromagnetic phase).

\section{Spontaneous symmetry breaking in the LMG model}\label{SSB}

According to Ehrenfest theorem, the equations of motion read
\begin{eqnarray}
	\frac{d  \vec S }{d t} & = & i  [H, \vec S] ,\\
	 \Rightarrow  \frac{d  S_x }{d t}&= &\frac{i}{N}  [S_z^2, S_x]  - i h [S_z, S_x],\nonumber \\
	& & = - \frac{1}{N}  (S_z S_y + S_y S_z)  + h  S_y,\nonumber \\
	& & = - \left(\frac{2 S_z}{N} - h\right) S_y;\label{meanx}\\
	 \Rightarrow  \frac{d  S_y }{d t} &=& \frac{i}{N}  [S_z^2, S_y]  - i h [S_z, S_y] ,\nonumber \\
	& & = \frac{1}{N}  (S_z S_x + S_x S_z ) - h  S_x,\nonumber \\
	& & = \left(\frac{2 S_z}{N} - h\right) S_x. \label{meany}
\end{eqnarray}
The last equality of Eq. (\ref{meanx}, \ref{meany}) holds under mean-field approximation $\langle S_{\alpha} S_{\beta} \rangle \approx \langle S_{\alpha} \rangle \langle S_{\beta} \rangle$. If we take second derivatives for Eq. (\ref{meanx}, \ref{meany}) with respect to time, the resulted mean-field dynamics implies a possible oscillation in the $x-y$ plane. The oscillation frequency $\omega = 2 \langle S_z \rangle / N - h \neq 0$ if $\langle S_z \rangle \neq Nh / 2$, in other words, $\omega \sim O(1 / N)$ if the ground-state energy is deviated from the classical minimum point.

\begin{figure}
	\centering
	\includegraphics[width=0.3\textwidth]{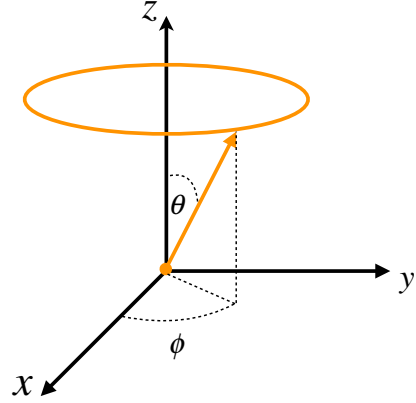}
	\caption{(Color online). Semiclassical picture of the infinite degeneracy of the isotropic LMG model in broken phase ($N \rightarrow \infty$).}
	\label{degeneracy}
\end{figure}

However, for a LMG model with rotational symmetry, every eigenstate has $\langle S_{x,y} \rangle = 0$ \cite{Dusuel2005} and thus the exact ground state cannot exhibit the above mean-field dynamics. One may consider that an external perturbation breaks the rotational symmetry in the $x-y$ plane such that $\langle S_{x,y} \rangle \neq 0$. The key point is to judge if this symmetry breaking happens spontaneously. Here we adopt the method presented by Bogoliubov \cite{Bogoliubov} to deal with the SSB of spin polarization in the LMG model, and we leave the discussion of time correlation function according to Watanabe's definition \cite{Watanabe2015} in section \uppercase\expandafter{\romannumeral5}. Assume we compute an order parameter $ \mathscr {O}(N, V)$ when both the number of particles $N$ and the external perturbation $V$ we are interested in are finite, then we take thermodynamic limit in this order: first $N \rightarrow \infty$, next $V \rightarrow 0$. If $\mathscr O$ vanishes in the thermodynamic limit, there is no SSB; while nonzero $\mathscr O$ indicates an SSB in the thermodynamic limit. By this definition, if we choose the in-plane polarization $m_{\vec n} = 2 \langle S_{\vec n} \rangle / N$ as the order parameter, where $\vec n = (\cos \phi, \sin \phi, 0)$ in the $x-y$ plane is the direction of an external magnetic field, the isotropic LMG model would spontaneously breaks the infinite degeneracy and select a mean-field ground state with  $m_{\vec n} \neq 0$ (see Fig. \ref{degeneracy}) in the thermodynamic limit. For example, under the symmetry-breaking potential $V = - g S_x$ where $g > 0$,  the perturbed Hamiltonian reads

\begin{equation}
	H = - \frac{1}{N}\left( S_x^2 + S_y^2 \right) - hS_z -gS_x,
\end{equation}

Assume we find the ground state $| \psi \rangle$ of the perturbed Hamiltonian, and then we can compute the ground state energy
\begin{equation}
	E_0 = - \frac{1}{N}\left( \langle S_x^2 \rangle + \langle S_y^2 \rangle \right) - h\langle S_z \rangle -g\langle S_x \rangle,
\end{equation}
and the order parameter is $m_x$. Then we take limit $N \rightarrow \infty$ to see the value of $E_0$ and $m_x$. When $N \rightarrow \infty$, the result should converge to the one computed through a mean-field (variational) approach. The mean-field wave function \cite{Dusuel2005}

\begin{equation}
	| \theta, \phi \rangle = \bigotimes_{l = 1}^N \left[ \cos  (\theta / 2) e^{-i\phi / 2} | \! \uparrow \rangle_l + \sin  (\theta / 2) e^{i\phi / 2} | \! \downarrow \rangle_l \right],
\end{equation}
where kets $| \! \! \uparrow \rangle_l$ and $| \! \! \downarrow \rangle_l$ are eigenstates of $\sigma_z^l$ with eigenvalues $+1$ and $-1$ separately. Besides, $| \theta, \phi \rangle$ is a coherent spin state with localized semiclassical polarization
\begin{equation}
	\langle \vec S \rangle = \frac{N}{2} (\sin \theta  \cos \phi, \sin  \theta \sin  \phi, \cos  \theta).
\end{equation}

The ground state is thus determined by minimizing the energy
\begin{equation}
	E_0 = -\frac{N}{4} ( \sin^2 \, \theta + 2h \cos  \theta + 2g \sin \theta   \cos  \phi ),
\end{equation}
One sets derivatives with respect to $\theta$ and $\phi$
 equal to zero and obtains
\begin{eqnarray}
	\sin  \theta_0 \, (\cos \theta_0 - h) & + & g \, \cos  \theta_0 \cos \phi_0 = 0, \label{derivative1}\\	
	\sin  \phi_0 & = & 0.
	\label{derivative2}
\end{eqnarray}

The semiclassical minimum point is at $(\theta_0, \phi_0)$. Then we take limit $g \rightarrow 0$, and solve Eq. (\ref{derivative1}, \ref{derivative2}) analytically with solution $\theta_0 = \text{arccos} \, h$ and $\phi_0 = 0$. Thus the order parameter $m_x = \sin  \theta_0 \neq 0$ in the thermodynamic limit, which confirms the existence of SSB of spin polarization in the LMG model.

Let's see how close is the energy splitting between a localized state and the exact ground state when $N$ is finite. 
When the ground state is localized by an infinitesimal symmetry-breaking potential, there exists an exponentially small energy splitting between mean-field ground states $\sim \text{exp}(-cN)$ \cite{Newman1977, Botet1983} (see appendix \ref{exp gap}), which can be seen as the energy gap opened by the tunneling between degenerate states. 
Based on Newman's calculation \cite{Newman1977}, some people believe energy splitting resulted from infinitesimal perturbation is exponentially small between the first excited state and the ground state \cite{Dusuel2005, Botet1983}, but they take for granted that both the first excited state and the exact ground state are mean-field localized, which is not true. To our best knowledge, no one have calculated the energy elevated by a mean-field ground state relative to the exact ground state before, which depends on the trial wave function for the mean-field ground state. If we adopt Newman's approximate mean-field ground state, the result is full of small d-matrix elements such that it is difficult for a precise estimation.

However, we can construct a trial wave function whose energy is O($1/N^2$) higher than the ground state
\begin{equation}
	| 0_{loc}^{\prime} \rangle \approx \left(1 -\frac{2}{N}\right)^{1/2} |0\rangle + \frac{1}{\sqrt{N}}|1\rangle + \frac{1}{\sqrt{N}}|2\rangle.
\end{equation}

Thus the energy gap $\Delta E = 2/N^2 \sim O(1/N^2)$ and the localized spin polarization $\langle S_x \rangle \sim O(\sqrt{N})$. According to the uncertainty relation between the lifetime and the energy per spin
\begin{equation}
	\frac{\Delta E}{N} \Delta t \geq \frac{\hbar}{2},
\end{equation}
the lifetime of this localized state is roughly $O(N^3)$.

Next we look into the size dependence for the oscillation frequency derived from Eq. (\ref{meanx}, \ref{meany}). Because $\omega \sim O(1 / N)$ vanish if $N \rightarrow \infty$, the selected mean-field ground state sticks there forever, which is similar to the well known superselection in the Ising model. But if $N$ is finite, would the spin polarization oscillate ``spontaneously"? To study the spontaneous property of the finite-size LMG model, in section \uppercase\expandafter{\romannumeral5} we will calculate the spin-spin correlation function in time domain. Since the exact spin polarization dynamics and the spin-spin time correlation function share common nature in their characteristic frequencies, we decide to derive the dynamics first in the next section, and then we go back to discuss the correlation function.

\section{Dynamics of the finite size LMG model}\label{dynamics}
Just like the case of a finite-size Ising model, although the SSB and the superselection are imperfect in a finite crystal, a tiny perturbation or fluctuation (which is also finite instead of infinitesimal) can still break the symmetry of the exact ground state and map it to a localized state which is close to the corresponding mean-field ground state.
In this section, we will solve the dynamics of the localized states of the LMG model with finite number of spins.

The exact equations of motion beyond mean-field approximation
\begin{eqnarray}
	\frac{d  S_x }{d t} & = & - \left(\frac{2 S_z}{N} - h\right) S_y - i \frac{1}{N} S_x; \\
	\frac{d  S_y }{d t} & = & \left(\frac{2 S_z}{N} - h\right) S_x - i \frac{1}{N} S_y.
\end{eqnarray}
In the derivation we use the commutation relation among the total spin operators
\begin{equation}
	[S_{\alpha}, S_{\beta}] = i \epsilon_{\alpha \beta \gamma} S_{\gamma}.
	\label{commutator}
\end{equation}

Notice that the differential equations are only defined being sandwiched by a pair of bra and ket. In order to solve the equations exactly, we have to separate two-spin operators $\langle S_z S_{x, y} \rangle$ into a product of one-spin operators $\langle S_z \rangle \langle S_{x, y}\rangle$. For the eigenstates when $\gamma = 1$, $\langle S_z S_{x, y} \rangle = \langle S_z \rangle \langle S_{x, y}\rangle$ exactly, so we consider to ``project" the original equations into the subspace of each eigenstate.

Here we use two eigenstate bases: $\{| m \rangle | m \in \mathbb Z^+\}$ in the representation of $S_z$, $\{| k \rangle | k \in \mathbb Z^+\}$ in the representation of $H$. The convention we adopt here is as follows: $\{m =  \text{0, 1, 2, ..., N}\}$ is a sequence arranged in descending order with respect to the corresponding eigenvalues of $S_z$, such that $\langle 0 | S_z | 0 \rangle = N / 2$ is the highest eigenvalue; $\{k = \text{N, N-1, ..., 0}\}$ is another sequence but arranged in ascending order with respect to the corresponding eigenvalues of $H$, such that $\langle 0 | H | 0 \rangle = E_0$ is the lowest eigenvalue. Since $[S_z, H] = 0$, $\{| m \rangle \}$ and $\{| k \rangle \}$ are the same basis up to a permutation $| m \rangle = P_{mk} | k \rangle$. The advantage to use these two bases is to keep track on the matrix elements of $S_{x,y}$ and the near-ground-state behavior simultaneously. Given a state $|\Psi \rangle = \sum_m c_m |m \rangle = \sum_k b_k |k \rangle$, we sandwich the whole differential equation between the bracket of this state, and it is necessary to compute expectation value of the two-spin operators. Take a look at $\langle S_z S_x \rangle$
\begin{eqnarray}
	\langle S_z S_x \rangle & = & \left(\sum_m c_m^{*} \langle m |\right) S_z S_x \left(\sum_{m^{\prime}} c_{m^{\prime}} | m^{\prime} \rangle\right) \nonumber \\
	& = & \sum_{m, m^{\prime}} M_k c_m^{*} c_{m^{\prime}} \langle m | S_x | m^{\prime} \rangle \nonumber \\
	& = & \sum_{k} M_k S_{x_k},
\end{eqnarray}
where
\begin{equation}
	S_{x_k} = c_m^{*} \left(c_{m + 1} S_{x_{m, m + 1}} + c_{m - 1} S_{x_{m, m - 1}}\right)
\end{equation}
denoting the projected solution contributed by $M_k = \langle k| S_z |k \rangle$. In addition, $S_{y_k}$ has the same form as $S_{x_k}$ by simply changing the subscript from $x$ to $y$. Thus the original differential equations reads
\begin{equation}
	\frac{d}{dt} \sum_k \left(\begin{array}{ccc}
					S_{x_k} \\
					S_{y_k}
				\end{array}\right) = \sum_k \left(\begin{array}{ccc}
									- \frac{i}{N} & h - \frac{2 M_k}{N} \\
									\frac{2 M_k}{N} - h & - \frac{i}{N}
									\end{array}\right) \left(\begin{array}{ccc}
														S_{x_k}\\
														S_{y_k}
													\end{array}\right),
\end{equation}
which can be decomposed into the sum of the projected equations
\begin{equation}
	\frac{d}{dt} \left(\begin{array}{ccc}
					S_{x_k} \\
					S_{y_k}
				\end{array}\right) = \left(\begin{array}{ccc}
									- \frac{i}{N} & h - \frac{2 M_k}{N} \\
									\frac{2 M_k}{N} - h & - \frac{i}{N}
									\end{array}\right) \left(\begin{array}{ccc}
														S_{x_k}\\
														S_{y_k}
													\end{array}\right),
\end{equation}

The projected solutions are
\begin{eqnarray}
	S_{x_k} & = & e^{-i\nu t} \left[S_{x_k}^0 \cos (\omega_k t) + S_{y_k}^0 \sin (\omega_k t)\right], \label{oscillationx}\\
	S_{y_k} & = & e^{-i\nu t} \left[S_{y_k}^0 \cos (\omega_k t) - S_{x_k}^0 \sin (\omega_k t)\right], \label{oscillationy}
\end{eqnarray}
where $S_{x_k}^0 = S_{x_k} (t = 0), S_{y_k}^0 = S_{y_k} (t = 0)$ are constants determined by the initial conditions. Notice that each projected solution is oscillating with two coupling intrinsic frequencies: $\nu = 1/N$ is universal for any subspace, while $\omega_k = h - 2 M_k / N$ depends on $k$. The complete solution is simply the superposition of every projected solution
\begin{eqnarray}
	\left\langle S_x (t) \right\rangle & = & \sum_{k = 0}^N S_{x_k} (t),\label{sx}\\
	\left\langle S_y (t) \right\rangle & = & \sum_{k = 0}^N S_{y_k} (t),\label{sy}.
\end{eqnarray}

As $N \rightarrow \infty$, both $\nu$ and $\omega_k$ goes to zero which coincides with the result given by Botet and Jullien\cite{Botet1983}, and thus no oscillation behavior in the thermodynamic limit. Recall $\nu$ comes from the commutator of the total spin operators and $\omega_0 \neq 0$ when the ground state energy is not exactly at the classical minimum point of the Hamiltonian, and therefore this kind of oscillation attributes to a purely quantum effect, i.e. a finite-size effect.

Here we discuss a special case when $Nh$ is also an integer. We show the parity of $N$ and $Nh$ classifies two modes of oscillation near the ground state with distinguished behaviors. Suppose through SSB we have already prepared a localized state which is really close to the exact ground state with $\langle S_{\vec n} (t = 0)\rangle \neq 0$, then we focus on its dynamics at $k = 0$ subspace.

\begin{figure}
	\centering
	\includegraphics[width=0.35\textwidth]{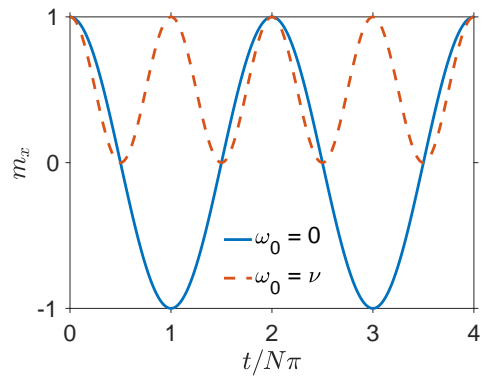}
	\caption{(Color online). Two modes of oscillation related to the parity of $N$ and $Nh$. $m_x=2 \langle S_x \rangle /N $ is the in-plane polarization along $x$ axis. $\omega_0 = 0$ specifies the round mode; $\omega_0 = \nu$ identifies the crescent mode.}
	\label{figure1}
\end{figure}

\begin{table}
	\centering
	\caption{Parity of $N$ and $Nh$ and the corresponding $\omega_0$. ``Deg." is the abbreviation of degeneracy and ``g.s." stands for the ground state at each case. }
	\begin{tabular}{c|c|c|c|c}
		\hline
		$N$ & $Nh$ & $M_0$ & deg. of g.s. & $\omega_0$\\
		\hline
		even & even & $Nh / 2$& nondeg. & 0\\
		\hline
		odd & odd & $Nh / 2$ & nondeg. & 0\\
		\hline
		even & odd & $(Nh \pm 1) / 2$ & deg. & $\nu$ \\
		\hline
		odd & even & $(Nh \pm 1) / 2$& deg. & $\nu$ \\
		\hline
	\end{tabular}
	\label{table1}
\end{table}

Two modes classified by the parity of $N$ and $Nh$ (see Table \ref{table1})

\begin{enumerate}
	\item Round mode: If $N$ and $Nh$ shares the same parity, then the ground state is unique, and $\omega_0 = 0$ so that the oscillation will be dominated by one frequency which is $\nu$. To emphasize, $\omega_0 = 0$ means the ground state energy is exactly at the classical minimum point, however there is still a nontrivial oscillation with frequency $\nu$ in the round mode, which is contributed by the commutator relation Eq. (\ref{commutator}). Thus $\nu$ reflects the quantum nature of spins.
	\item Crescent mode: If parities of $N$ and $Nh$ are different, then there exist a two-fold degeneracy at the ground state, and $\omega_0 = \nu$. The coupling between $\nu$ and $\omega_0$ generates a oscillation with frequency equal to $2\nu$.
\end{enumerate}

Fig. \ref{figure1} shows the two modes of $\langle S_x (t) \rangle$ with the initial conditions $\langle S_x (t = 0) \rangle = N/2$ and $\langle S_y (t = 0) \rangle = 0$.
Here we define a polarization $ m_x (t)= 2 \langle S_x (t) \rangle/N$
For $\langle S_y (t) \rangle$, it is the same as $\langle S_x (t) \rangle$ up to a $\pi/2$ phase difference, and thus the physical picture here is the expectation value of the total spin is precessing along the $z$ axis. Notice that both modes have single frequency, but the crescent mode oscillates twice faster than the round mode if $N$ is the same. In addition, another interesting difference is that the round mode can rotate a full circle, while the crescent mode bounces back and forth restricted in half a circle.

However, if the external perturbation is too small, it may be hard to observe the crescent mode. The exact ground states of crescent mode is twofold degenerate (see Table \ref{table1}), and we label them with $| \! \! \uparrow \rangle$ and $| \! \! \downarrow \rangle$ in the representation of $S_z$. From degenerate perturbation theory, we can easily compute the energy gap opened by the symmetry-breaking perturbation $V = - g S_x$. The perturbed Hamiltonian in the subspace spanned by $| \! \! \uparrow \rangle$ and $|\! \! \downarrow\rangle$ is
\begin{equation}
	V = -g S_{x_{\uparrow, \downarrow}}
	\left(
		\begin{array}{cc}
			0 & 1 \\
			1 & 0
		\end{array}
	\right).
\end{equation}

The eigenvalues are $\epsilon_{\pm} = \pm g S_{x_{\uparrow, \downarrow}}$ where $S_{x_{ \uparrow, \downarrow}} = \langle \uparrow| S_x   |\! \! \downarrow\rangle$, and the corresponding eigenstates are $(|\! \! \uparrow\rangle \pm |\! \! \downarrow\rangle)/\sqrt{2}$. Since the energy difference vanishes at $g = 0$, states $(|\! \! \uparrow\rangle \pm |\! \! \downarrow\rangle)/\sqrt{2}$ with nonzero $\langle S_x \rangle$ have exactly the same energy as the ground state, so it seems like the crescent mode under this superposition state should last forever with frequency $2\nu = 2/N$. However, since states $(|\! \! \uparrow\rangle \pm |\! \! \downarrow\rangle)/\sqrt{2}$ are still the exact ground state, $\langle e^{iHt} S_x e^{iHt} \rangle = S_{x_{\uparrow, \downarrow}}$ is just a constant without any time dependence.

In general, $Nh$ is not an integer and thus there are more than one frequency components of the oscillation. Through Fourier analysis, each $S_{x_k,y_k}$ has two frequency components $\nu \pm \omega_k$. Furthermore, if we look at a localized state near the ground state, since $|c_0|^2 \approx 1$ and $|c_k|^2 \approx 0$ for $k \neq 0$, then $\langle S_{x, y}(t) \rangle \approx S_{x_0, y_0}(t)$ and $\nu \pm \omega_0$ will dominate the oscillation behavior.

\section{Correlation function and relation with time crystals}
In this section we will calculate the correlation function according to the definition given by Watanabe and et al.\cite{Watanabe2015} in order to answer if time-translational symmetry would be broken ``spontaneously" in the finite-size LMG model. The correlation function corresponding to the order parameter $\hat m_x = 2 \hat S_x / N$ is
\begin{equation}
	\langle \hat m_x(t) \hat m_x(0) \rangle = \langle e^{i \hat H t} \hat m_x e^{-i \hat H t} \hat m_x \rangle = f_N(t).
\end{equation}
The correlation function is sandwiched by the ground state. The condition that we believe time-translation symmetry is broken is read $\text{lim}_{N \rightarrow \infty} \langle e^{i \hat H t} \hat m_x e^{-i \hat H t} \hat m_x \rangle = \text{lim}_{N \rightarrow \infty} f_N(t) = F(t)$, where $F(t)$ is a periodic function in time, and such condition defines a time crystal. However, for any physical system in the equilibrium state, time crystal behavior is impossible which has been proven by the no-go theorem\cite{Bruno2013b, Watanabe2015}. Therefore, we decide to loose the condition somehow and define a condition for ``effective" time crystal behavior in a finite-size system: although $F(t)$ has no time dependence, if $f_N(t)$ shows nontrivial periodic oscillation when $N$ is finite, and the frequency component $\nu$ of $f_N(t)$ is a function of $N$ and has form
\begin{equation}
	\nu(N) = \nu_0 \phi(N) = \nu_0 (N^{-1} + a_2 N^{-2} +... + a_l N^{-p}),
\end{equation}
where $\phi(N)$ is expanded near the infinity, and the lifetime of a oscillating state should be longer than $O(N^p)$. If we do a variable substitution such that $u \rightarrow 1/N$, then $\phi(u)$ is a polynomial of finite order $p$. The lower the $p$, the slower the frequency decreases, and the property of long range order in this finite system is better.

We now can go back to calculate $f_N(t)$ when $N$ is large but finite.
By convention according to section \uppercase\expandafter{\romannumeral4}, suppose the ground state $|0\rangle$ in $H$ representation is the state $|m_0\rangle$ in $S_z$ representation, then we compute the correlation function explicitly
\begin{eqnarray}
	\lefteqn{ \frac{N^2}{4}f_N(t) = \langle 0 | e^{i \hat H t} S_x e^{-i \hat H t} S_x | 0 \rangle}\nonumber \\
	& = & e^{i E_0 t} \langle m_0 | \sum_{m_k^{\prime}} \sum_{m_k}  e^{-i E_k t} S_{x_{m_k^{\prime},m_k}} S_{x_{m_k,m_0}} | m_k^{\prime} \rangle\nonumber \\
	& = & |S_{x_{m_0,m_1}}|^2 \left( e^{i (E_0 - E_1) t} + e^{i (E_0 - E_2) t} \right),
	\label{correlation}
\end{eqnarray}
where $|E_{1,2} - E_0| = 1/N \pm (2M_0/N -h)$ actually equals $\nu \pm \omega_0$, and since $m_{1, 2} = m_0 \pm 1$ are defined in the $S_z$ representation, so $S_{x_{m_0,m_1}} \neq 0$. This result is surprisingly simple, which shows the ground state coupling with the first and the second excited states and produces two oscillating modes. But notice that $S_x = (S_+ + S_-)/2$ can only raise or lower the angular momentum quantum number by 1 for $S_z$ 's eigenstates, therefore this coupling in Eq. (\ref{correlation}) is expected. Someone may argue that this computation is operated the subspace of $S=N/2$ sector instead of in the direct product space of $N$ spin $1/2$, and the result maybe in doubt. However, Eq. (\ref{correlation}) is indeed correct, and the proof is as follows. The whole direct product space can be decomposed into the direct sum of different total angular momentum sectors
\begin{equation}
	\bigotimes_{n = 1}^N\mathscr{H_n}_{\frac{1}{2}} =
	\left\{
	\begin{array}{ll}
		\bigoplus_{n = 1}^{\frac{N}{2}}\mathscr{H}(S=n), &\text{if N is even,} \\
		\bigoplus_{n = 0}^{\frac{N-1}{2}}\mathscr{H}(S=n + \frac{1}{2}), &\text{if N is odd.}
	\end{array}
	\right.
\end{equation}

Since the ground state is certainly located in $S=N/2$ sector due to ferromagnetic interaction, the projection of ground state in other $S\neq N/2$ sectors is zero vector $\vec0$, and thus when operator $S_x = S_x^{S=N/2} \bigoplus S_x^{S\neq N/2}$ hits the ground state $|0\rangle = |0\rangle^{S=N/2} \bigoplus \vec0^{S\neq N/2}$, there is no chance for the result being mapped into other $S\neq N/2$ sectors. Therefore it is safe to do calculation of correlation function only in the $S=N/2$ sector.
Referring to our discussion of the properties of $\nu \pm \omega_0$ in section \uppercase\expandafter{\romannumeral3} and \uppercase\expandafter{\romannumeral4}, the frequency components of $f_N(t)$ decrease as $1/N$, and the lifetime of the oscillation near ground state increases as $N^3$. Thus we may link the oscillating dynamics in the finite-size LMG model to effective time crystal behavior.

\section{Numerical simulation}

\begin{figure}
	\centering
	\includegraphics[width=0.23\textwidth]{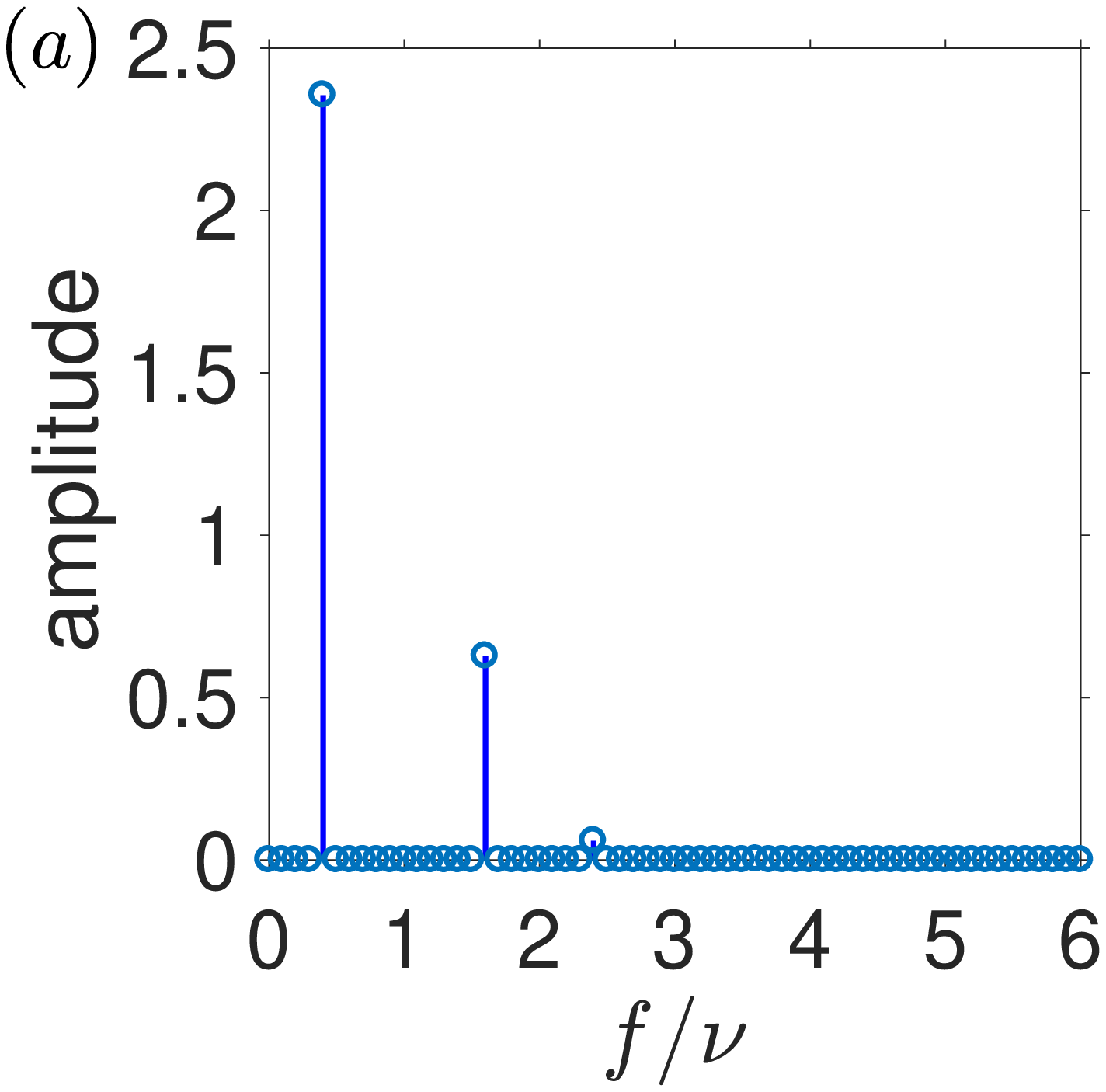}
	\includegraphics[width=0.217\textwidth]{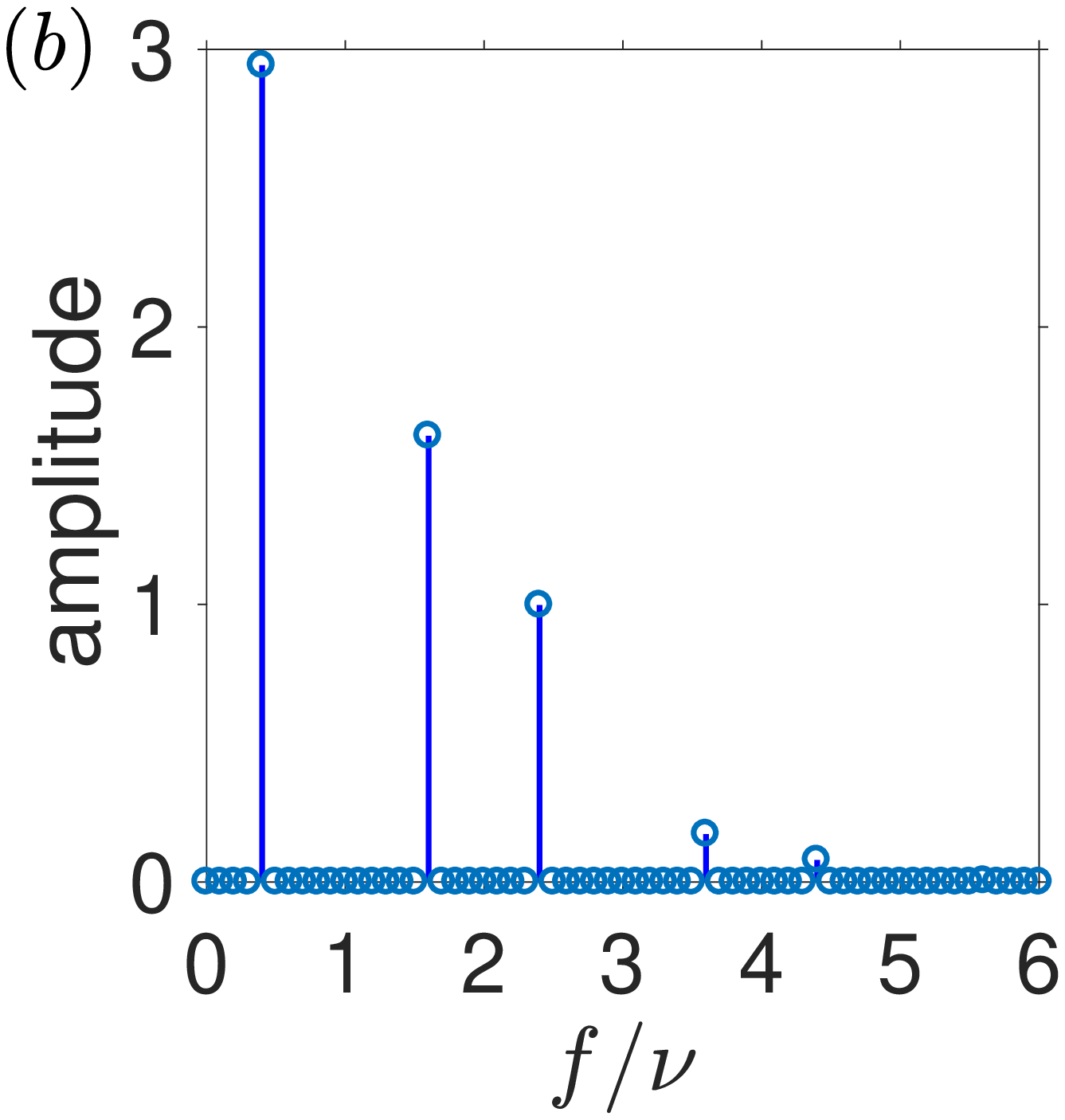}
	\caption{(Color online). Frequency spectrums in a $N = 100$ and $h = 0.714$ system with different magnitude of symmetric-breaking potential $V = - g S_x$. (a) $g = 1 \times 10^{-4}$. The energy difference between this localized state and the exact ground state $\Delta E = 6.60 \times 10^{-4}$. (b) $g = 1 \times 10^{-3}$, with $\Delta E = 5.10 \times 10^{-3}$.}
	\label{perturbation}
\end{figure}

\begin{figure}
	\centering
	\includegraphics[width=0.243\textwidth]{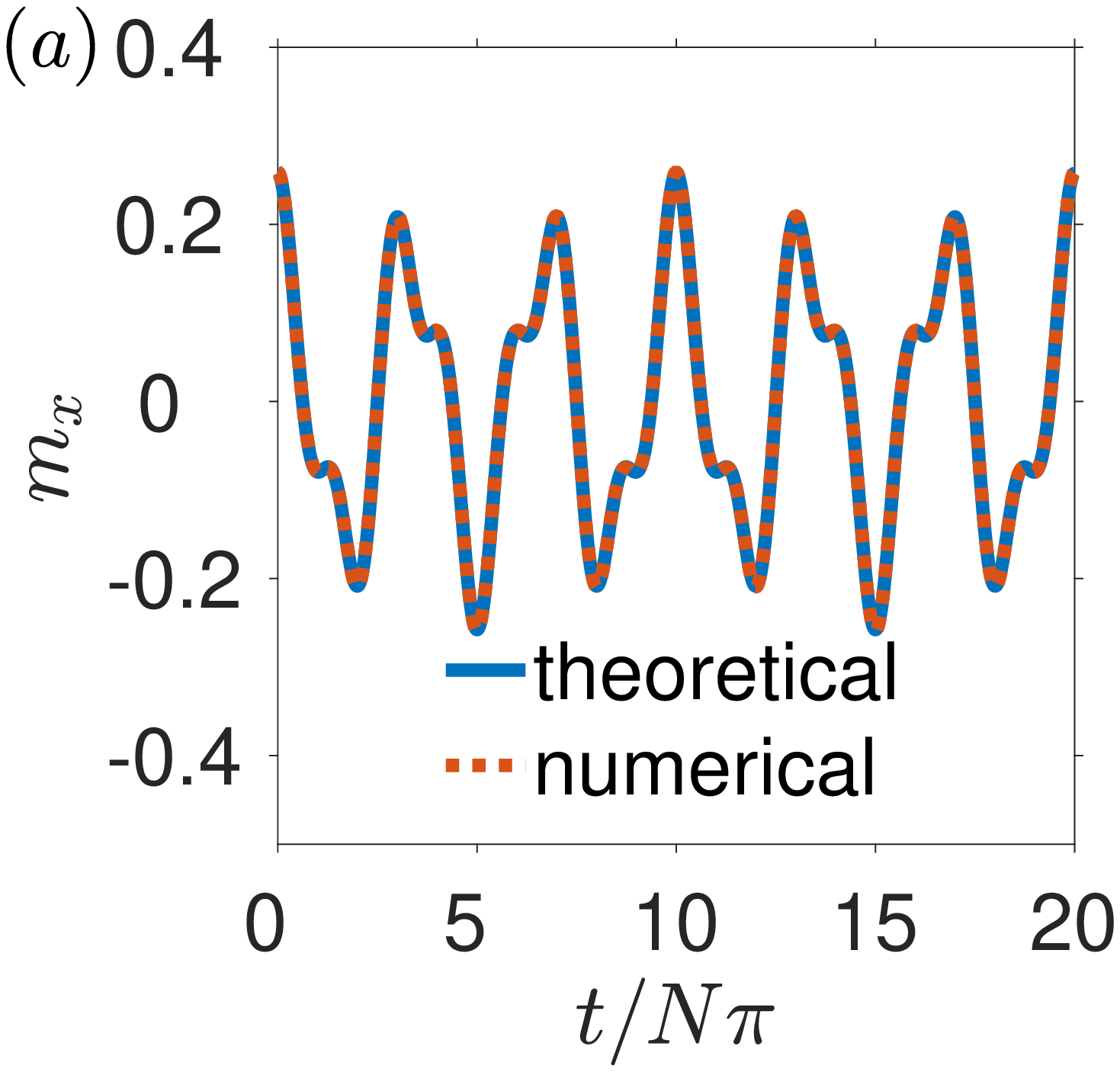}
	\includegraphics[width=0.23\textwidth]{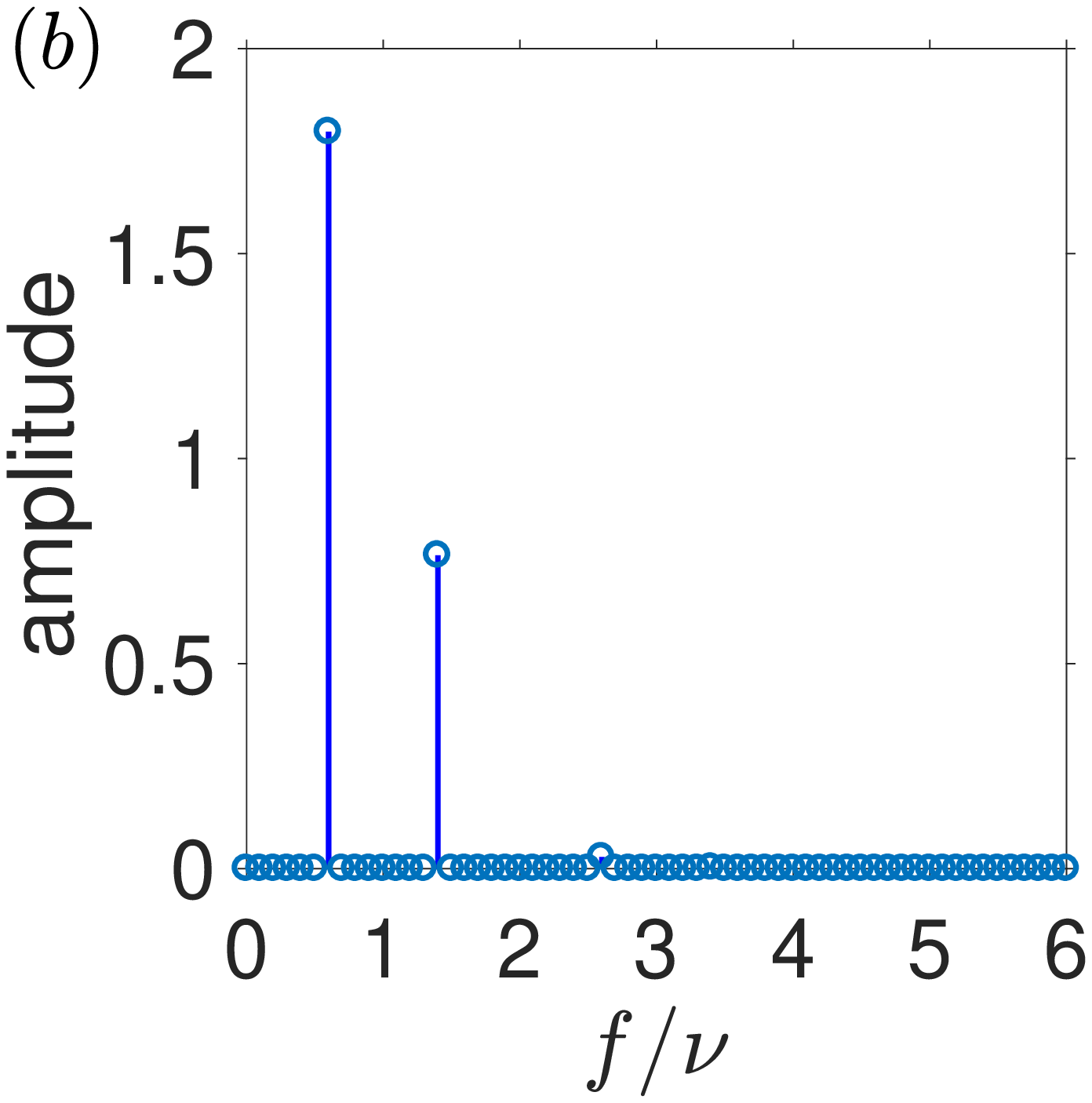}
	\caption{(Color online). (a) An example of  $m_x (t) $ oscillation with $N = 100$, $h = 0.716$. The localized state is obtained by bringing into an instantaneous symmetry-breaking field $V = - g S_x$ at $t = 0$, where $g = 1 / N^2$. (b) The frequency spectrum of the oscillation in (a).}
	\label{kiss}
\end{figure}

\begin{figure}
	\centering
	\includegraphics[width=0.23\textwidth]{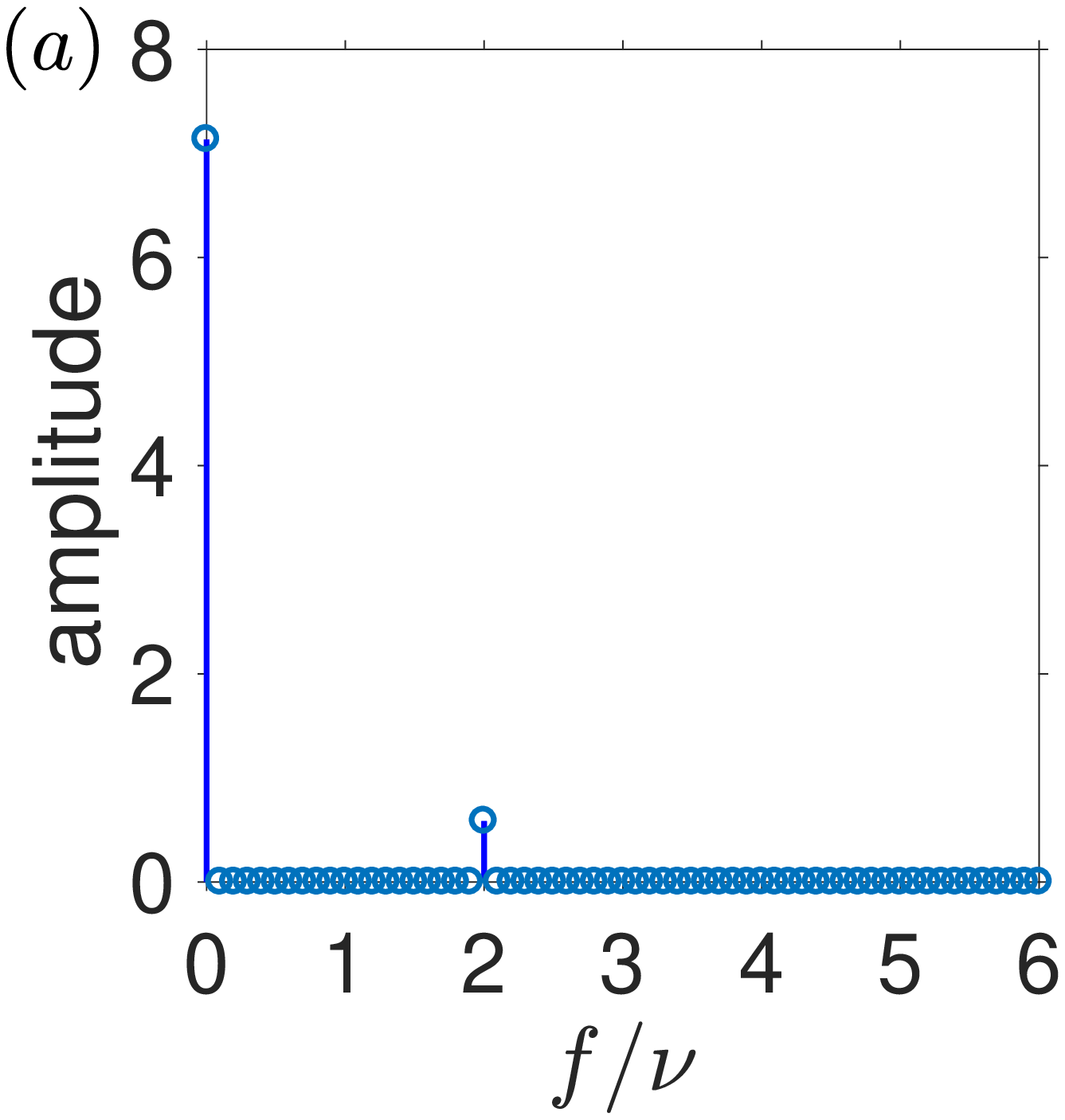}
	\includegraphics[width=0.232\textwidth]{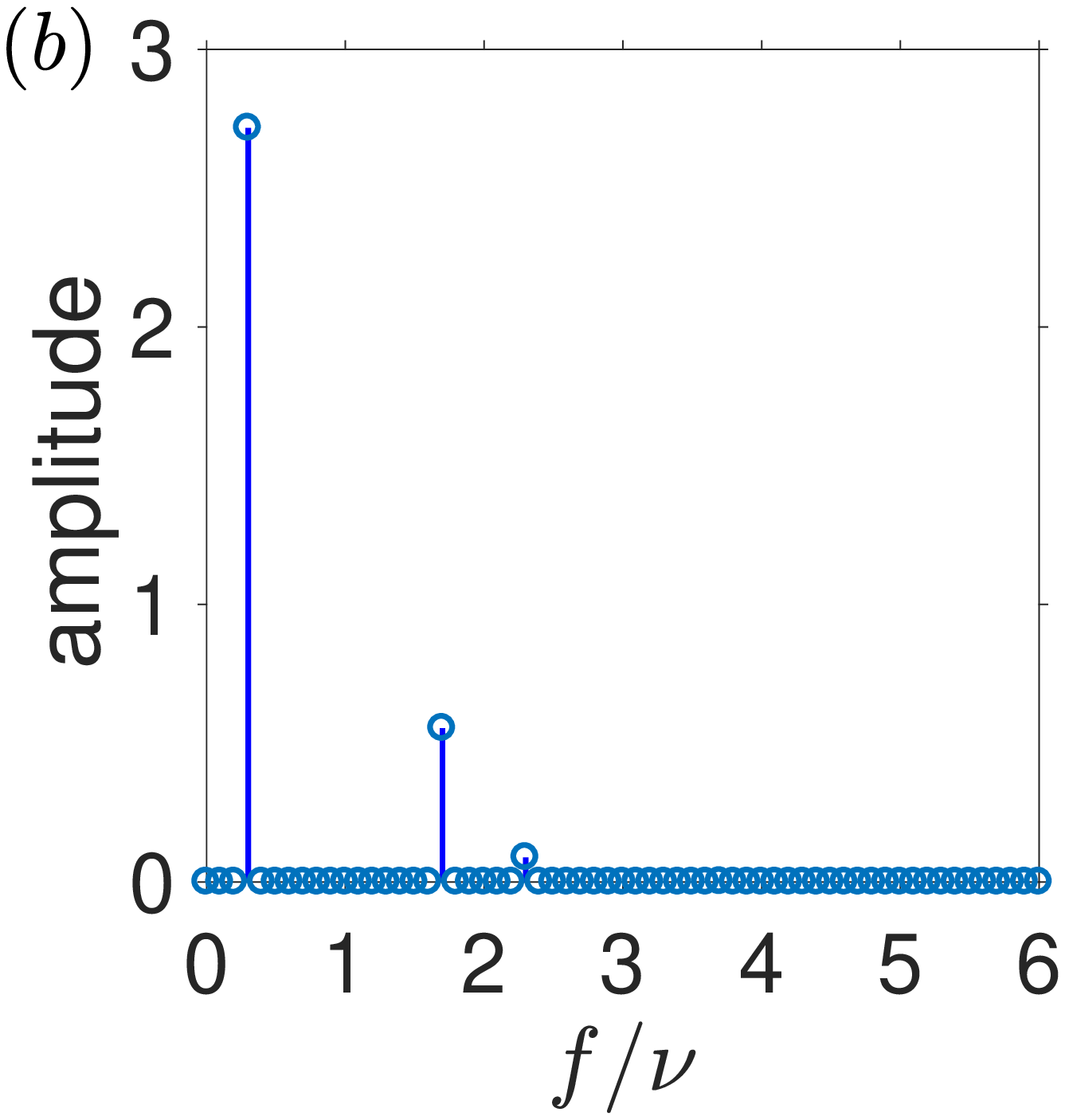}
	\includegraphics[width=0.236\textwidth]{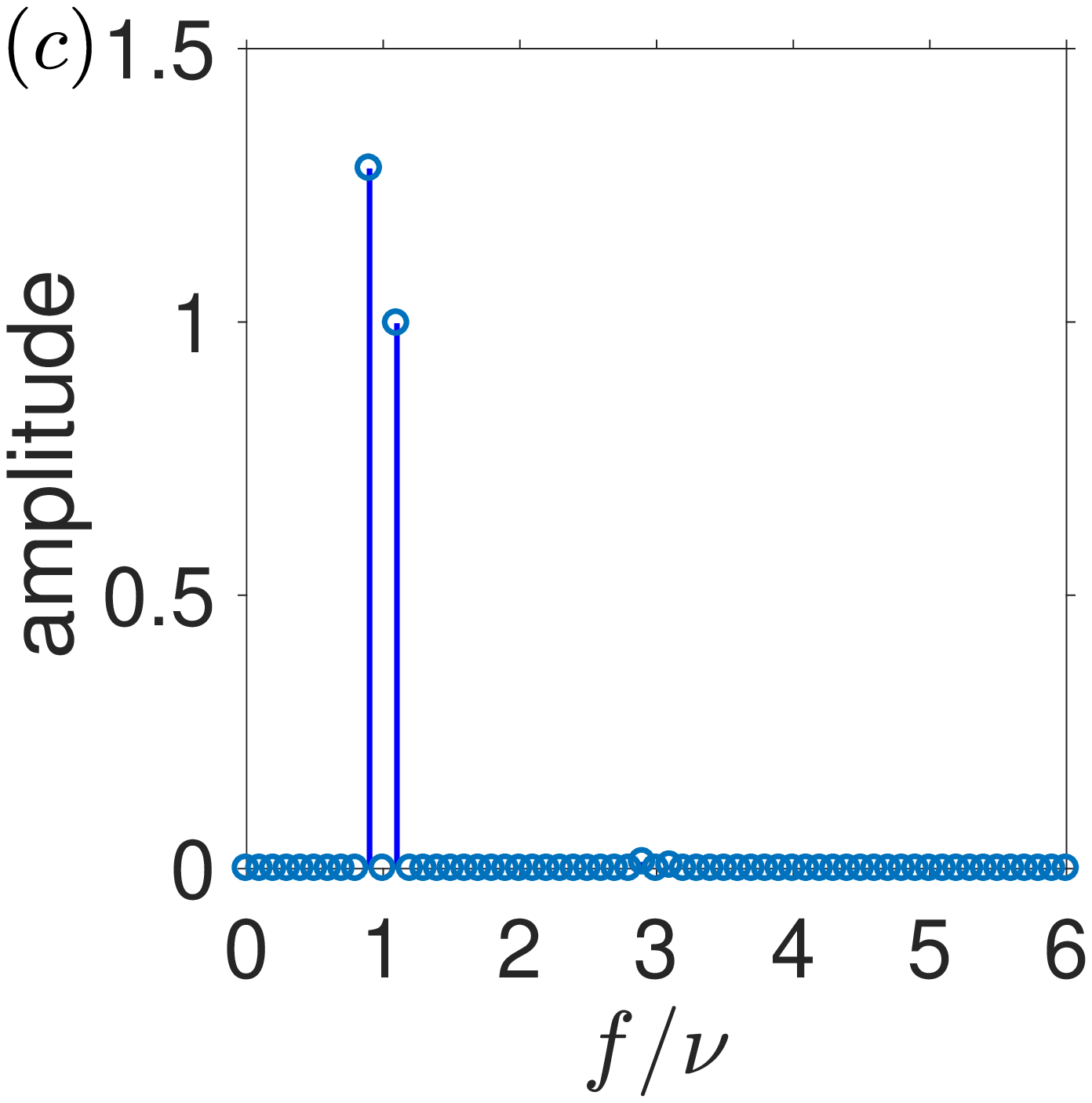}
	\includegraphics[width=0.236\textwidth]{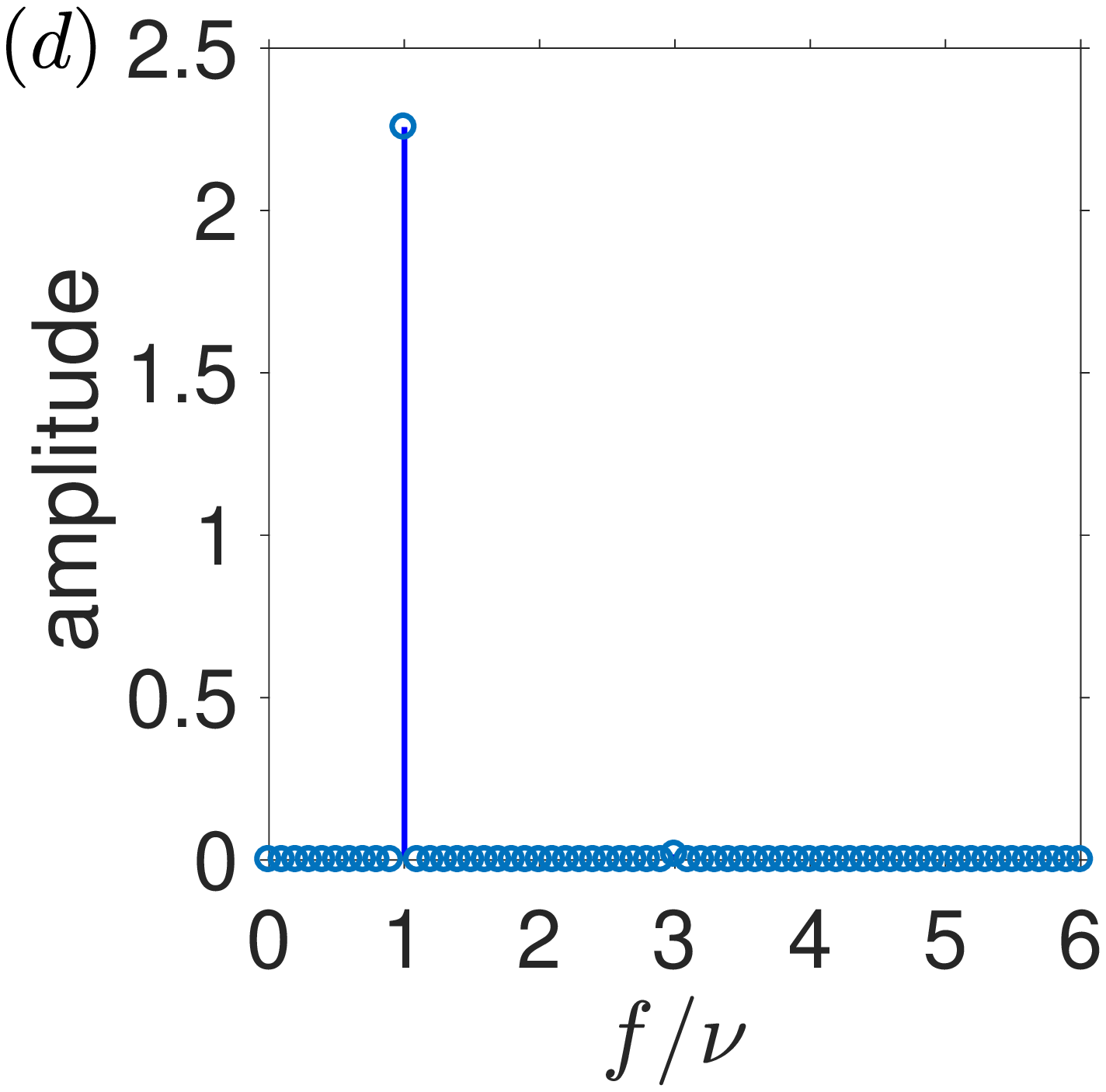}
	\caption{(Color online). The above frequency spectrums are computed from a $N = 100$ system with different $h$. (a) $h = 0.710$; since $Nh = 71$ is odd, (a) characterizes a crescent mode. (b) $h = 0.713$. (c) $h = 0.719$. (d) $h = 0.720$; both $N$ and $Nh$ are even, so (d) represents a round mode. See Table \ref{table1}.}
	\label{close}
\end{figure}

\begin{figure}
	\centering
	\includegraphics[width=0.39\textwidth]{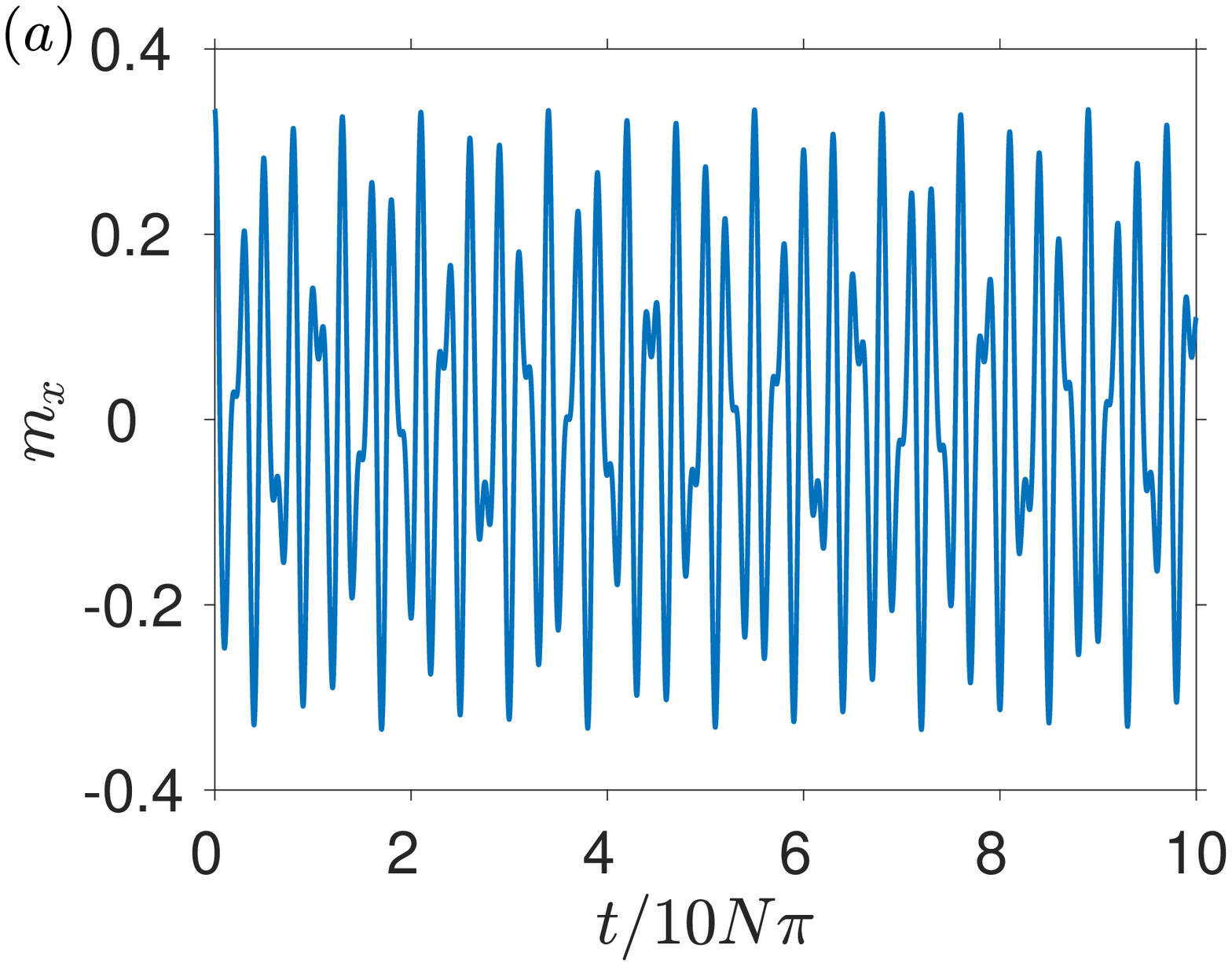}
	\includegraphics[width=0.37\textwidth]{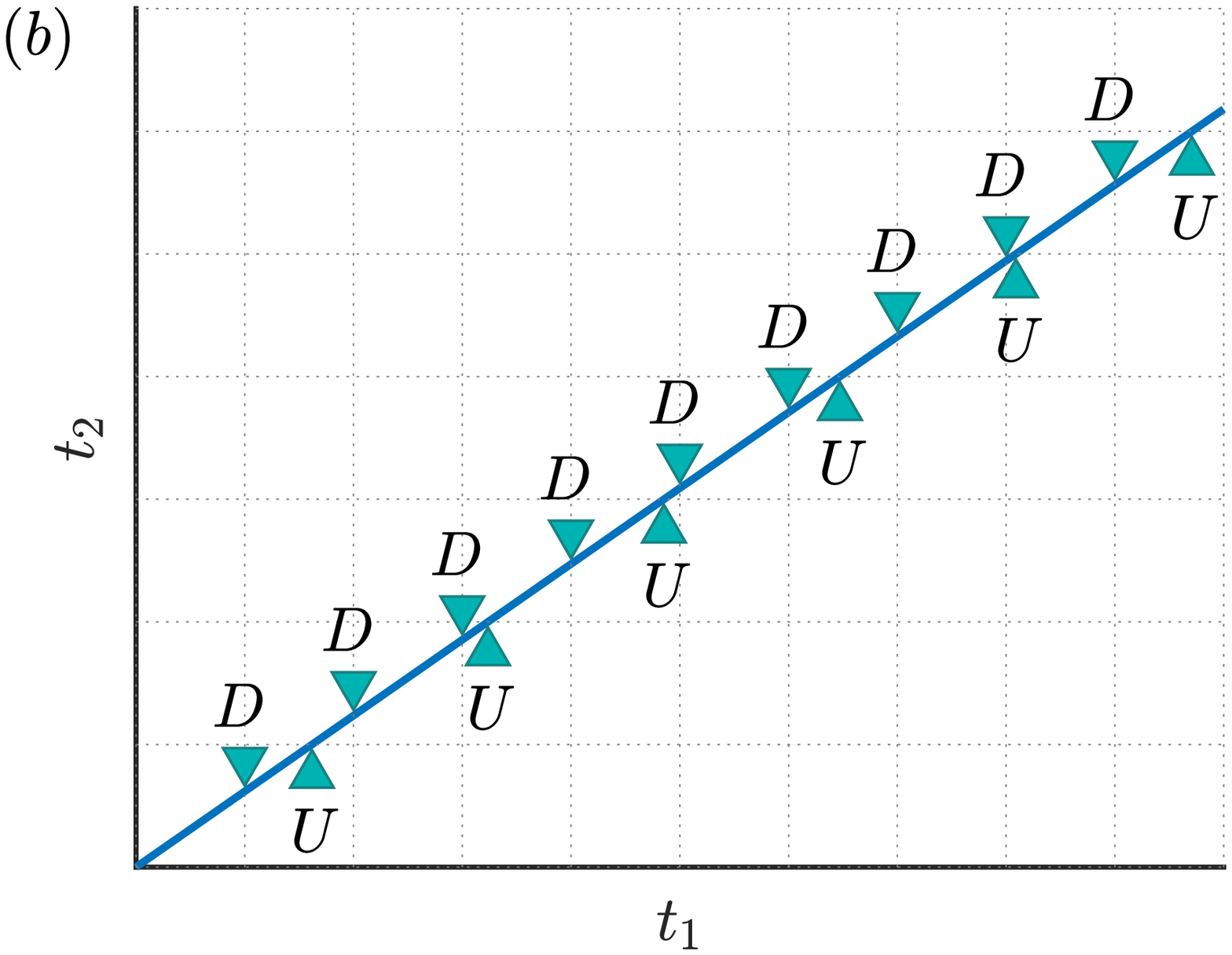}
	\caption{(Color online).(a) An example of aperiodic oscillation ($S_{x_0}$ component) in $N= 100$ and $h = \nu (52 + \omega_0)$ system, where $\nu =  1/100$ and $\omega_0 = \nu (\sqrt{5} - 1)/(\sqrt{5} + 3)$. Two frequency components for this oscillation: $f_1 = \nu - \omega_0 $ and $f_2 = \nu + \omega_0 $. $f_1 / f_2 = (\sqrt{5} -1)/2$ is the golden ratio, and thus it is related to the Fibonacci quasicrystal \cite{2016arXiv160808220B,timequasicrystal}. (b) Generating 1D quasicrystals by the intersection method. An irrational slice, whose slope is equal to the golden ratio $(\sqrt{5} -1)/2$, cuts through a 2D lattice. We label $U$ (up triangles) or $D$ (down triangles) when horizontal or vertical lines are crossed.}
	\label{quasicrystal}
\end{figure}

The numerical simulation is built upon $S = N/2$ subspace (ferromagnetic phase)  because we are only interested in the low-energy states. The dimension of this subspace is $N + 1$, which is much smaller than the dimension of the total direct-product space $2^N$, so we can compute the exact evolution of quite a large system. To select a localized state, we add an instantaneous symmetry-breaking potential into the Hamiltonian

\begin{equation}
	H = H^0 + H^{\prime},\label{H}
\end{equation}

where $H^0$ is the Hamiltonian in Eq. (\ref{h0}) and $H^{\prime} = g S_{\vec n}$ represents an instantaneous symmetry-breaking magnetic field along $\vec n$ at $t = 0$. Diagonalize $H$ in Eq. (\ref{H}) numerically and choose its ground state as the approximate localized state $| 0_{loc}^{\prime} \rangle$, we should see the in-plane oscillation behavior predicted by Eq. (\ref{sx}, \ref{sy}) as $| 0_{loc}^{\prime} \rangle$ evolves under $H^0$. Since $H^0$ is time-independent, the time evolution is simply calculated as $e^{i H^0 t} | 0_{loc}^{\prime} \rangle$. In our numerical simulation we set $H^{\prime} = g S_x$ with $N = 100$ and $g = 1/100^2$ as an example. The results are shown in Fig. \ref{perturbation}-\ref{quasicrystal}, which agree with the theoretical prediction.

In fact, this symmetry-breaking potential cannot be too large. Fig. \ref{perturbation}. (b) shows more higher-frequency modes are excited which means the energy difference is relatively large and thus the lifetime of this state may be too short to demonstrate the low-frequency dynamics.

Here we use the first three terms in Eq. (\ref{sx}, \ref{sy}) as our theoretical prediction of the in-plane oscillation since projected solutions with smaller $k$ play more important role in the whole solution if the initial state is really close to the ground state

\begin{equation}
	\left\langle S_x (t) \right\rangle \approx \sum_{k = 0}^3 S_{x_k} (t), \;\;\;
	\left\langle S_y (t) \right\rangle \approx \sum_{k = 0}^3 S_{y_k} (t).\label{approx1}
\end{equation}
	
From Fig. \ref{kiss} (a), we see that, in the case of $N = 100 $ and $g = 1 / 100^2$, the theoretical prediction of the wave form almost matches the numerical simulation exactly. The spectrum in Fig. \ref{kiss} (b) is dominated by two frequencies $0.6 \nu$ and $1.4 \nu$ which are exactly the two frequencies $|\nu \pm \omega_0|$ contributed by $S_{x_0}$, and the third tiny peak is at $2.6 \nu$, which is one of $|\nu \pm \omega_1|$ provided by $S_{x_1}$. This result confirms that $S_{x_0}$ reveals the most important component of $\langle S_x (t) \rangle$, which makes Eq. (\ref{approx1}) a really good approximation.

Recall in section \uppercase\expandafter{\romannumeral4}, we specify two single-frequency modes according to the parities of $N$ and $Nh$, and they are also substantiated by the numerical simulation, which are Fig. \ref{close} (a) a crescent mode with frequency $2\nu$ and Fig. \ref{close} (d) a round mode with frequency $\nu$. In addition, as shown in Fig. \ref{close}, the lower the frequency locates the higher it peaks. An physically intuitive interpretation is that those lower-frequency modes have lower energies so they must be excited first. When the lowest two frequencies are close, one finds their peaks also approach equally high, and it can be explained by their near-degenerate energies.

One of the most fascinated nature of the finite-size LMG model is there are two intrinsic frequencies at its ground state, so it is natural to consider the possibility to realize a 1D quasicrystal in time domain.
Quasicrystals in time domain, or so-called ``time quasicrystals" were proposed by Li and et al. in space-time crystals of trapped ions \cite{Li2012}.
Recently, it was investigated in open systems with and without Floquet driving \cite{timequasicrystal,timequasicrystal1,2017arXiv171010087G}.

Generally speaking, quasicrystal is defined as a system that is quasiperiodic and has crystallographically forbidden rotational symmetry simultaneously, but this requires at least two dimensions.
For time series which are one-dimensional, quasiperiodicity with higher-dimensional counterpart in the Penrose tiling can be viewed as quasicrystals.

In our finite-size LMG model, if $h$ is irrational, the ratio of the two intrinsic frequencies will also be irrational and thus ensures a quasiperiodic oscillation in the time domain. What is more, if the irrational frequency ratio is related to the Penrose tiling, then we may think of this quasiperiodic oscillation an effective time quasicrystal. For instance, the golden ratio $(\sqrt{5} -1)/2$ defines a Fibonacci quasicrystal \cite{timequasicrystal} (see Fig. \ref{quasicrystal}). 
Here is one way to construct any ratio $\kappa \in (0,1)$ by tuning $h$. Referring to our previous result in section \uppercase\expandafter{\romannumeral4}, the frequency ratio reads $\kappa = (\nu - \omega_0)/(\nu + \omega_0)$, and for a given $\kappa$ we solve for $\omega_0(\kappa) = \nu (1 - \kappa)/(1 + \kappa)$. According to Eq. (\ref{groundm}, \ref{oscillationx}, \ref{oscillationy}), $\omega_0$ is also a function of $h$, so eventually we can write $h(\kappa)$ explicitly
\begin{equation}
	h =
	\left \{
	\begin{array}{ll}
		\nu \left(2\zeta + \frac{1 - \kappa}{1 + \kappa}\right), &\text{when N is even,}\\
		&\text{$\zeta = 0, 1, ..., (N-2)/2$;}\\
		\nu \left(2\eta + 1 + \frac{1 - \kappa}{1 + \kappa}\right), &\text{when N is odd,}\\
		&\text{$\eta = 0, 1, ..., (N-3)/2$.}
	\end{array}
	\right.
\end{equation}

In other words, by controlling $h$ which is the external magnetic field along $z$ axis, one may get quasicrystal behavior spontaneously. 

\section{Discussion and Conclusion}
Recently, Syrwid and et al. proposed that time crystal behavior could be found in the excited eigenstates of Wilzeck's model \cite{Syrwid2017}. There are some similarities between Syrwid's results and ours, and we make a short comparison. For both cases, thermal influence has to be sufficiently week to preserve the localized state. In Wilczek's model, lifetime of a state is defined by contrast of density-density correlation function. Numerical calculation shows the time crystal behavior lifetime of an excited state is linearly increased with $N$. In our model, we defined lifetime directly from the uncertainty relation between energy and time. We give a theoretical trial wave function near ground state whose lifetime increases as $N^3$.

Before conclusion, we briefly discuss how to realize the LMG model and test the symmetry-breaking dynamics in experiment.
In order to observe the dynamics of the LMG model near the ground state, the number of spins should be much larger than $1$.  As
shown in numerical results, when $N=100$, the exact dynamics of the system is very close the theoretical predictions. Besides, the dynamics
of the LMG model will be greatly affected by the number of spins. Therefore, in experiments, the ability to precisely control or maintain the
the spins is necessary. In superconducting circuit systems, more than $10$ superconducting qubits have been coupled with the same resonator
\cite{2017arXiv170310302S}, which may be used for realizing the LMG model and testing the predictions in this paper. Besides, as some of us
discussed in Ref. \cite{MaYue2017}, NV centers in optically trapped nanodiamond may also be used for simulating the LMG model.

In conclusion, we studied the nearly spontaneous dynamics of the finite size LMG model. We found that the ground state of the finite size LMG model can be localized
with tiny perturbation. The localized state would oscillate with two different intrinsic frequencies, both of which are $\sim O(1/N)$.
Take the trial wave function in section \uppercase\expandafter{\romannumeral3} as an example, its lifetime is in proportional to $N^3$, so the dynamics of the localized state would last for long enough time when $N$ is large.
This phenomenon closely relates to the original definition of the quantum time crystal. Moreover, the two frequencies are usually irrational. Therefore, the dynamics of the
localized state is not periodic, which can be connected to a quantum time quasicrystal.

\acknowledgments
We thank the helpful discussions with Peng Zhang, Dongling Deng, Julien Vidal, and Paul J. Steinhardt.
Z.-Q. Y. is supported by the National Natural Science Foundation of China Grant 61771278, 61435007 and 11574176. T. L. acknowledges the support from NSF under Grant No. 1555035-PHY. Y. H. acknowledges support from the Education Program for Talented Students of Xi'an Jiaotong University.

\appendix

\section{The Exponentially Small Gap Between Mean-Field Ground States} \label{exp gap}

Here we will give a brief derivation, referring to Newman and Schulman \cite{Newman1977}, of the exponentially small gap between mean-field ground states when an infinitesimal symmetry-breaking potential is turned on near the thermodynamic limit ($N \rightarrow \infty$). Recall the LMG Hamiltonian
\begin{equation}
	H = -\frac{1}{N} (S_x^2 + \gamma S_y^2) - h S_z, \label{h00}
\end{equation}
For simplicity, we set $\gamma = 0$ so that the mean-field ground states in broken phase ($0<h<1$) have two-fold degeneracy instead of infinite degeneracy which is hard to compute. 

We turn on the symmetry-breaking perturbation $V = - g S_x$, and divide the perturbed Hamiltonian by the spin number so that we define the average energy per spin
\begin{equation}
	H_N(g) = -\frac{S_x^2}{N^2} - \frac{h S_z}{N} - \frac{g S_x}{N}, \label{h1}
\end{equation}

To study ground state energy $E_N^0(g)$ near $g = 0$, we introduce a set of states $| m, \theta \rangle$ ($m = -\frac{N}{2}, -\frac{N}{2} + 1, ..., \frac{N}{2}$) which are eigenstates of $S_{\theta} = e^{-i \theta S_y} S_z e^{i \theta S_y}$, $S_{\theta} | m, \theta \rangle = m | m, \theta \rangle $, and try to find $\theta$ such that $| m, \theta \rangle$ is approximately an eigenstate of $H_N(g)$ when $N$ is very large. In other words, the off diagonal matrix elements of $H_N(g)$ in the basis $\{ | m, \theta \rangle \}$ approaches zero when $N$ is large.
\begin{figure}
	\centering
	\includegraphics[width=0.30\textwidth]{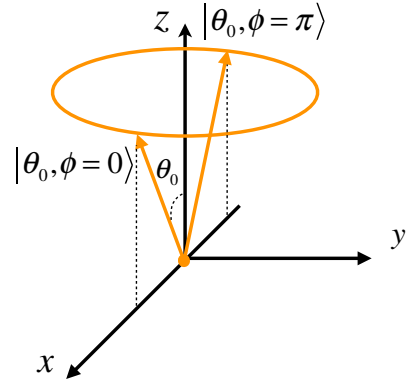}
	\caption{(Color online). Two-fold degeneracy when $\gamma = 0$.}
	\label{two fold}
\end{figure}
\begin{eqnarray}
	\lefteqn{\langle m^{\prime}, \theta | H_N(g) | m, \theta \rangle = } \nonumber \\
	& - & \delta_{m^{\prime}, m} \{(1/N^2) m^2 \, \sin^2\theta + h (m/N) \cos\theta + g (m/N) \sin\theta\nonumber \\
	& + &  (1/N^2) \cos^2\theta[(N/2)(N/2 + 1) - m^2]\} \nonumber \\
	& + & [2S_z(m^{\prime}, m)/N]\{\sin\theta \, \cos\theta[(m + m^{\prime})/N]  \nonumber \\
	& - & h \, \sin\theta + g \, \cos\theta\} \nonumber \\
	& - & (1/4N^2) \cos^2\theta (S_+^2 + S_-^2)(m^{\prime}, m).
\end{eqnarray}

where $A(m^{\prime}, m) = \langle m^{\prime}, \theta | A | m, \theta \rangle$. If $m, m^{\prime} = N/2 - O(1)$ as $N \rightarrow \infty$, off diagonal elements in Eq. (1.3) vanish to order $1/N$ if we demand that
\begin{equation}
	\sin\theta \, \cos\theta - h \, \sin\theta + g \, \cos\theta = 0, \label{off diagonal 0}
\end{equation}


For $h \neq 0$ the two solutions of \ref{off diagonal 0}, $\theta_1$ and $\theta_2$, are mean-field locations minimizing the energy under perturbation, we thus consider $| m, \theta_1 \rangle$ and $| m, \theta_2 \rangle$ as approximately the two-fold degenerate ground states of $H_N(h)$. If $|g| \ll 1$, then $\theta_1 + \theta_2 \approx 0$. According to degenerate perturbation theory, we look at the subspace spanned by $| N/2, \theta_0 \rangle$ and $| N/2, - \theta_0 \rangle$ for small $|g|$, where $\theta_0 = \text{arccos}\,h \in (0, \pi / 2]$. 

The $2\times2$ matrix approximation to the subspace of $H_N(g)$ is as follows
\begin{eqnarray}
	\lefteqn{\left[ -\frac{1 + h^2}{4} + O\left(\frac{1}{N}\right) \right]
	\left(
		\begin{array}{cc}
			1 & 0 \\
			0 & 1
		\end{array}
	\right)} \nonumber \\
	& - &
	\left(
		\begin{array}{cc}
			0 & \alpha \\
			\alpha & 0
		\end{array}
	\right)
	- \frac{g}{2}
	\left(
		\begin{array}{cc}
			\sin\theta_0 & 0 \\
			0 & -\sin\theta_0
		\end{array}
	\right),	
\end{eqnarray}

where $\alpha = \langle N/2, \theta_0 | H_N(g) | N/2, -\theta_0 \rangle$. As $N \rightarrow \infty$, $| N/2, \pm \theta_0 \rangle$ are approximate ground states of $H_N(g)$ up to an $O(1/N)$ correction, and we can express $| N/2, -\theta_0 \rangle = e^{i (2 \theta_0) S_y} | N/2, \theta_0 \rangle$ using rotation matrix, so $\alpha$ can be computed and is exponentially small with respect to $N$:
\begin{eqnarray}
	\lefteqn{\alpha \approx H_N(0) d^{\left(\frac{N}{2}\right)}_{\frac{N}{2}\frac{N}{2}}(2 \theta_0)}\nonumber \\
	& \propto & (\cos\theta_0)^N = \text{exp}[-(-\text{ln}h)N] = \text{exp}[-c(h)N]
\end{eqnarray}

where $d^{(j)}_{m m^{\prime}}(\beta)$ is an element of Wigner's d-matrix, $c(h) = -\text{ln}h > 0$ for $0<h<1$. Thus the energy splitting between $E_N^0$ and $E_N^1$ near $h = 0$ is approximated by the eigenvalues of
\begin{equation}
	W = -
	\left(
		\begin{array}{cc}
			0 & \alpha \\
			\alpha & 0
		\end{array}
	\right)
	-\frac{g}{2}
	\left(
		\begin{array}{cc}
			(1 - h^2)^{1/2} & 0 \\
			0 & -(1 - h^2)^{1/2}
		\end{array}
	\right).
\end{equation}

The eigenvalues of $W$ are $\epsilon_{\pm} = \pm [\alpha^2 + g^2(1 - h^2)/4]^{1/2}$, so the energy gap at $g = 0$ is of size $2\alpha$, which is also exponentially small with $N$.

%

\begin{thebibliography}{47}%
\makeatletter
\providecommand \@ifxundefined [1]{%
 \@ifx{#1\undefined}
}%
\providecommand \@ifnum [1]{%
 \ifnum #1\expandafter \@firstoftwo
 \else \expandafter \@secondoftwo
 \fi
}%
\providecommand \@ifx [1]{%
 \ifx #1\expandafter \@firstoftwo
 \else \expandafter \@secondoftwo
 \fi
}%
\providecommand \natexlab [1]{#1}%
\providecommand \enquote  [1]{``#1''}%
\providecommand \bibnamefont  [1]{#1}%
\providecommand \bibfnamefont [1]{#1}%
\providecommand \citenamefont [1]{#1}%
\providecommand \href@noop [0]{\@secondoftwo}%
\providecommand \href [0]{\begingroup \@sanitize@url \@href}%
\providecommand \@href[1]{\@@startlink{#1}\@@href}%
\providecommand \@@href[1]{\endgroup#1\@@endlink}%
\providecommand \@sanitize@url [0]{\catcode `\\12\catcode `\$12\catcode
  `\&12\catcode `\#12\catcode `\^12\catcode `\_12\catcode `\%12\relax}%
\providecommand \@@startlink[1]{}%
\providecommand \@@endlink[0]{}%
\providecommand \url  [0]{\begingroup\@sanitize@url \@url }%
\providecommand \@url [1]{\endgroup\@href {#1}{\urlprefix }}%
\providecommand \urlprefix  [0]{URL }%
\providecommand \Eprint [0]{\href }%
\providecommand \doibase [0]{http://dx.doi.org/}%
\providecommand \selectlanguage [0]{\@gobble}%
\providecommand \bibinfo  [0]{\@secondoftwo}%
\providecommand \bibfield  [0]{\@secondoftwo}%
\providecommand \translation [1]{[#1]}%
\providecommand \BibitemOpen [0]{}%
\providecommand \bibitemStop [0]{}%
\providecommand \bibitemNoStop [0]{.\EOS\space}%
\providecommand \EOS [0]{\spacefactor3000\relax}%
\providecommand \BibitemShut  [1]{\csname bibitem#1\endcsname}%
\let\auto@bib@innerbib\@empty
\bibitem [{\citenamefont {Anderson}(1963)}]{Anderson}%
  \BibitemOpen
  \bibfield  {author} {\bibinfo {author} {\bibfnamefont {P.~W.}\ \bibnamefont
  {Anderson}},\ }\href {\doibase 10.1103/PhysRev.130.439} {\bibfield  {journal}
  {\bibinfo  {journal} {Phys. Rev.}\ }\textbf {\bibinfo {volume} {130}},\
  \bibinfo {pages} {439} (\bibinfo {year} {1963})}\BibitemShut {NoStop}%
\bibitem [{\citenamefont {Higgs}(1964)}]{Higgs}%
  \BibitemOpen
  \bibfield  {author} {\bibinfo {author} {\bibfnamefont {P.~W.}\ \bibnamefont
  {Higgs}},\ }\href {\doibase 10.1103/PhysRevLett.13.508} {\bibfield  {journal}
  {\bibinfo  {journal} {Phys. Rev. Lett.}\ }\textbf {\bibinfo {volume} {13}},\
  \bibinfo {pages} {508} (\bibinfo {year} {1964})}\BibitemShut {NoStop}%
\bibitem [{\citenamefont {Wilczek}(2012)}]{Wilczek2012}%
  \BibitemOpen
  \bibfield  {author} {\bibinfo {author} {\bibfnamefont {F.}~\bibnamefont
  {Wilczek}},\ }\href {\doibase 10.1103/PhysRevLett.109.160401} {\bibfield
  {journal} {\bibinfo  {journal} {Phys. Rev. Lett.}\ }\textbf {\bibinfo
  {volume} {109}},\ \bibinfo {pages} {160401} (\bibinfo {year}
  {2012})}\BibitemShut {NoStop}%
\bibitem [{\citenamefont {Shapere}\ and\ \citenamefont
  {Wilczek}(2012)}]{Shapere2012}%
  \BibitemOpen
  \bibfield  {author} {\bibinfo {author} {\bibfnamefont {A.}~\bibnamefont
  {Shapere}}\ and\ \bibinfo {author} {\bibfnamefont {F.}~\bibnamefont
  {Wilczek}},\ }\href {\doibase 10.1103/PhysRevLett.109.160402} {\bibfield
  {journal} {\bibinfo  {journal} {Phys Rev Lett}\ }\textbf {\bibinfo {volume}
  {109}},\ \bibinfo {pages} {160402} (\bibinfo {year} {2012})}\BibitemShut
  {NoStop}%
\bibitem [{\citenamefont {Li}\ \emph {et~al.}(2012)\citenamefont {Li},
  \citenamefont {Gong}, \citenamefont {Yin}, \citenamefont {Quan},
  \citenamefont {Yin}, \citenamefont {Zhang}, \citenamefont {Duan},\ and\
  \citenamefont {Zhang}}]{Li2012}%
  \BibitemOpen
  \bibfield  {author} {\bibinfo {author} {\bibfnamefont {T.}~\bibnamefont
  {Li}}, \bibinfo {author} {\bibfnamefont {Z.-X.}\ \bibnamefont {Gong}},
  \bibinfo {author} {\bibfnamefont {Z.-Q.}\ \bibnamefont {Yin}}, \bibinfo
  {author} {\bibfnamefont {H.~T.}\ \bibnamefont {Quan}}, \bibinfo {author}
  {\bibfnamefont {X.}~\bibnamefont {Yin}}, \bibinfo {author} {\bibfnamefont
  {P.}~\bibnamefont {Zhang}}, \bibinfo {author} {\bibfnamefont {L.-M.}\
  \bibnamefont {Duan}}, \ and\ \bibinfo {author} {\bibfnamefont
  {X.}~\bibnamefont {Zhang}},\ }\href {\doibase 10.1103/PhysRevLett.109.163001}
  {\bibfield  {journal} {\bibinfo  {journal} {Phys Rev Lett}\ }\textbf
  {\bibinfo {volume} {109}},\ \bibinfo {pages} {163001} (\bibinfo {year}
  {2012})}\BibitemShut {NoStop}%
\bibitem [{\citenamefont {Bruno}(2013{\natexlab{a}})}]{Bruno2013a}%
  \BibitemOpen
  \bibfield  {author} {\bibinfo {author} {\bibfnamefont {P.}~\bibnamefont
  {Bruno}},\ }\href {\doibase 10.1103/PhysRevLett.111.029301} {\bibfield
  {journal} {\bibinfo  {journal} {Phys Rev Lett}\ }\textbf {\bibinfo {volume}
  {111}},\ \bibinfo {pages} {029301} (\bibinfo {year}
  {2013}{\natexlab{a}})}\BibitemShut {NoStop}%
\bibitem [{\citenamefont {Bruno}(2013{\natexlab{b}})}]{Bruno2013b}%
  \BibitemOpen
  \bibfield  {author} {\bibinfo {author} {\bibfnamefont {P.}~\bibnamefont
  {Bruno}},\ }\href {\doibase 10.1103/PhysRevLett.111.070402} {\bibfield
  {journal} {\bibinfo  {journal} {Phys Rev Lett}\ }\textbf {\bibinfo {volume}
  {111}},\ \bibinfo {pages} {070402} (\bibinfo {year}
  {2013}{\natexlab{b}})}\BibitemShut {NoStop}%
\bibitem [{\citenamefont {Wilczek}(2013{\natexlab{a}})}]{Wilczek2013a}%
  \BibitemOpen
  \bibfield  {author} {\bibinfo {author} {\bibfnamefont {F.}~\bibnamefont
  {Wilczek}},\ }\href {\doibase 10.1103/PhysRevLett.110.118902} {\bibfield
  {journal} {\bibinfo  {journal} {Phys. Rev. Lett.}\ }\textbf {\bibinfo
  {volume} {110}},\ \bibinfo {pages} {118902} (\bibinfo {year}
  {2013}{\natexlab{a}})}\BibitemShut {NoStop}%
\bibitem [{\citenamefont {{Li}}\ \emph {et~al.}(2012)\citenamefont {{Li}},
  \citenamefont {{Gong}}, \citenamefont {{Yin}}, \citenamefont {{Quan}},
  \citenamefont {{Yin}}, \citenamefont {{Zhang}}, \citenamefont {{Duan}},\ and\
  \citenamefont {{Zhang}}}]{arxiv1212.6959v2}%
  \BibitemOpen
  \bibfield  {author} {\bibinfo {author} {\bibfnamefont {T.}~\bibnamefont
  {{Li}}}, \bibinfo {author} {\bibfnamefont {Z.-X.}\ \bibnamefont {{Gong}}},
  \bibinfo {author} {\bibfnamefont {Z.-Q.}\ \bibnamefont {{Yin}}}, \bibinfo
  {author} {\bibfnamefont {H.~T.}\ \bibnamefont {{Quan}}}, \bibinfo {author}
  {\bibfnamefont {X.}~\bibnamefont {{Yin}}}, \bibinfo {author} {\bibfnamefont
  {P.}~\bibnamefont {{Zhang}}}, \bibinfo {author} {\bibfnamefont {L.-M.}\
  \bibnamefont {{Duan}}}, \ and\ \bibinfo {author} {\bibfnamefont
  {X.}~\bibnamefont {{Zhang}}},\ }\href@noop {} {\bibfield  {journal} {\bibinfo
   {journal} {ArXiv e-prints}\ } (\bibinfo {year} {2012})},\ \Eprint
  {http://arxiv.org/abs/1212.6959} {arXiv:1212.6959 [quant-ph]} \BibitemShut
  {NoStop}%
\bibitem [{\citenamefont {Watanabe}\ and\ \citenamefont
  {Oshikawa}(2015)}]{Watanabe2015}%
  \BibitemOpen
  \bibfield  {author} {\bibinfo {author} {\bibfnamefont {H.}~\bibnamefont
  {Watanabe}}\ and\ \bibinfo {author} {\bibfnamefont {M.}~\bibnamefont
  {Oshikawa}},\ }\href {\doibase 10.1103/PhysRevLett.114.251603} {\bibfield
  {journal} {\bibinfo  {journal} {Phys Rev Lett}\ }\textbf {\bibinfo {volume}
  {114}},\ \bibinfo {pages} {251603} (\bibinfo {year} {2015})}\BibitemShut
  {NoStop}%
\bibitem [{\citenamefont {Wilczek}(2013{\natexlab{b}})}]{Wilczek2013}%
  \BibitemOpen
  \bibfield  {author} {\bibinfo {author} {\bibfnamefont {F.}~\bibnamefont
  {Wilczek}},\ }\href {\doibase 10.1103/PhysRevLett.111.250402} {\bibfield
  {journal} {\bibinfo  {journal} {Phys. Rev. Lett.}\ }\textbf {\bibinfo
  {volume} {111}},\ \bibinfo {pages} {250402} (\bibinfo {year}
  {2013}{\natexlab{b}})}\BibitemShut {NoStop}%
\bibitem [{\citenamefont {Sacha}(2015)}]{Sacha2015}%
  \BibitemOpen
  \bibfield  {author} {\bibinfo {author} {\bibfnamefont {K.}~\bibnamefont
  {Sacha}},\ }\href {\doibase 10.1103/PhysRevA.91.033617} {\bibfield  {journal}
  {\bibinfo  {journal} {Phys. Rev. A}\ }\textbf {\bibinfo {volume} {91}},\
  \bibinfo {pages} {033617} (\bibinfo {year} {2015})}\BibitemShut {NoStop}%
\bibitem [{\citenamefont {Khemani}\ \emph {et~al.}(2016)\citenamefont
  {Khemani}, \citenamefont {Lazarides}, \citenamefont {Moessner},\ and\
  \citenamefont {Sondhi}}]{PhysRevLett.116.250401}%
  \BibitemOpen
  \bibfield  {author} {\bibinfo {author} {\bibfnamefont {V.}~\bibnamefont
  {Khemani}}, \bibinfo {author} {\bibfnamefont {A.}~\bibnamefont {Lazarides}},
  \bibinfo {author} {\bibfnamefont {R.}~\bibnamefont {Moessner}}, \ and\
  \bibinfo {author} {\bibfnamefont {S.~L.}\ \bibnamefont {Sondhi}},\ }\href
  {\doibase 10.1103/PhysRevLett.116.250401} {\bibfield  {journal} {\bibinfo
  {journal} {Phys. Rev. Lett.}\ }\textbf {\bibinfo {volume} {116}},\ \bibinfo
  {pages} {250401} (\bibinfo {year} {2016})}\BibitemShut {NoStop}%
\bibitem [{\citenamefont {Else}\ \emph {et~al.}(2016)\citenamefont {Else},
  \citenamefont {Bauer},\ and\ \citenamefont {Nayak}}]{PhysRevLett.117.090402}%
  \BibitemOpen
  \bibfield  {author} {\bibinfo {author} {\bibfnamefont {D.~V.}\ \bibnamefont
  {Else}}, \bibinfo {author} {\bibfnamefont {B.}~\bibnamefont {Bauer}}, \ and\
  \bibinfo {author} {\bibfnamefont {C.}~\bibnamefont {Nayak}},\ }\href
  {\doibase 10.1103/PhysRevLett.117.090402} {\bibfield  {journal} {\bibinfo
  {journal} {Phys. Rev. Lett.}\ }\textbf {\bibinfo {volume} {117}},\ \bibinfo
  {pages} {090402} (\bibinfo {year} {2016})}\BibitemShut {NoStop}%
\bibitem [{\citenamefont {Yao}\ \emph {et~al.}(2017)\citenamefont {Yao},
  \citenamefont {Potter}, \citenamefont {Potirniche},\ and\ \citenamefont
  {Vishwanath}}]{Yao2017}%
  \BibitemOpen
  \bibfield  {author} {\bibinfo {author} {\bibfnamefont {N.~Y.}\ \bibnamefont
  {Yao}}, \bibinfo {author} {\bibfnamefont {A.~C.}\ \bibnamefont {Potter}},
  \bibinfo {author} {\bibfnamefont {I.-D.}\ \bibnamefont {Potirniche}}, \ and\
  \bibinfo {author} {\bibfnamefont {A.}~\bibnamefont {Vishwanath}},\ }\href
  {\doibase 10.1103/PhysRevLett.118.030401} {\bibfield  {journal} {\bibinfo
  {journal} {Phys. Rev. Lett.}\ }\textbf {\bibinfo {volume} {118}},\ \bibinfo
  {pages} {030401} (\bibinfo {year} {2017})}\BibitemShut {NoStop}%
\bibitem [{\citenamefont {Else}\ \emph {et~al.}(2017)\citenamefont {Else},
  \citenamefont {Bauer},\ and\ \citenamefont {Nayak}}]{PhysRevX.7.011026}%
  \BibitemOpen
  \bibfield  {author} {\bibinfo {author} {\bibfnamefont {D.~V.}\ \bibnamefont
  {Else}}, \bibinfo {author} {\bibfnamefont {B.}~\bibnamefont {Bauer}}, \ and\
  \bibinfo {author} {\bibfnamefont {C.}~\bibnamefont {Nayak}},\ }\href
  {\doibase 10.1103/PhysRevX.7.011026} {\bibfield  {journal} {\bibinfo
  {journal} {Phys. Rev. X}\ }\textbf {\bibinfo {volume} {7}},\ \bibinfo {pages}
  {011026} (\bibinfo {year} {2017})}\BibitemShut {NoStop}%
\bibitem [{\citenamefont {{Zhang}}\ \emph {et~al.}(2017)\citenamefont
  {{Zhang}}, \citenamefont {{Hess}}, \citenamefont {{Kyprianidis}},
  \citenamefont {{Becker}}, \citenamefont {{Lee}}, \citenamefont {{Smith}},
  \citenamefont {{Pagano}}, \citenamefont {{Potirniche}}, \citenamefont
  {{Potter}}, \citenamefont {{Vishwanath}}, \citenamefont {{Yao}},\ and\
  \citenamefont {{Monroe}}}]{Zhang2017}%
  \BibitemOpen
  \bibfield  {author} {\bibinfo {author} {\bibfnamefont {J.}~\bibnamefont
  {{Zhang}}}, \bibinfo {author} {\bibfnamefont {P.~W.}\ \bibnamefont {{Hess}}},
  \bibinfo {author} {\bibfnamefont {A.}~\bibnamefont {{Kyprianidis}}}, \bibinfo
  {author} {\bibfnamefont {P.}~\bibnamefont {{Becker}}}, \bibinfo {author}
  {\bibfnamefont {A.}~\bibnamefont {{Lee}}}, \bibinfo {author} {\bibfnamefont
  {J.}~\bibnamefont {{Smith}}}, \bibinfo {author} {\bibfnamefont
  {G.}~\bibnamefont {{Pagano}}}, \bibinfo {author} {\bibfnamefont {I.-D.}\
  \bibnamefont {{Potirniche}}}, \bibinfo {author} {\bibfnamefont {A.~C.}\
  \bibnamefont {{Potter}}}, \bibinfo {author} {\bibfnamefont {A.}~\bibnamefont
  {{Vishwanath}}}, \bibinfo {author} {\bibfnamefont {N.~Y.}\ \bibnamefont
  {{Yao}}}, \ and\ \bibinfo {author} {\bibfnamefont {C.}~\bibnamefont
  {{Monroe}}},\ }\href {\doibase 10.1038/nature21413} {\bibfield  {journal}
  {\bibinfo  {journal} {Nature}\ }\textbf {\bibinfo {volume} {543}},\ \bibinfo
  {pages} {217} (\bibinfo {year} {2017})},\ \Eprint
  {http://arxiv.org/abs/1609.08684} {arXiv:1609.08684 [quant-ph]} \BibitemShut
  {NoStop}%
\bibitem [{\citenamefont {{Choi}}\ \emph {et~al.}(2017)\citenamefont {{Choi}},
  \citenamefont {{Choi}}, \citenamefont {{Landig}}, \citenamefont {{Kucsko}},
  \citenamefont {{Zhou}}, \citenamefont {{Isoya}}, \citenamefont {{Jelezko}},
  \citenamefont {{Onoda}}, \citenamefont {{Sumiya}}, \citenamefont {{Khemani}},
  \citenamefont {{von Keyserlingk}}, \citenamefont {{Yao}}, \citenamefont
  {{Demler}},\ and\ \citenamefont {{Lukin}}}]{Choi2017}%
  \BibitemOpen
  \bibfield  {author} {\bibinfo {author} {\bibfnamefont {S.}~\bibnamefont
  {{Choi}}}, \bibinfo {author} {\bibfnamefont {J.}~\bibnamefont {{Choi}}},
  \bibinfo {author} {\bibfnamefont {R.}~\bibnamefont {{Landig}}}, \bibinfo
  {author} {\bibfnamefont {G.}~\bibnamefont {{Kucsko}}}, \bibinfo {author}
  {\bibfnamefont {H.}~\bibnamefont {{Zhou}}}, \bibinfo {author} {\bibfnamefont
  {J.}~\bibnamefont {{Isoya}}}, \bibinfo {author} {\bibfnamefont
  {F.}~\bibnamefont {{Jelezko}}}, \bibinfo {author} {\bibfnamefont
  {S.}~\bibnamefont {{Onoda}}}, \bibinfo {author} {\bibfnamefont
  {H.}~\bibnamefont {{Sumiya}}}, \bibinfo {author} {\bibfnamefont
  {V.}~\bibnamefont {{Khemani}}}, \bibinfo {author} {\bibfnamefont
  {C.}~\bibnamefont {{von Keyserlingk}}}, \bibinfo {author} {\bibfnamefont
  {N.~Y.}\ \bibnamefont {{Yao}}}, \bibinfo {author} {\bibfnamefont
  {E.}~\bibnamefont {{Demler}}}, \ and\ \bibinfo {author} {\bibfnamefont
  {M.~D.}\ \bibnamefont {{Lukin}}},\ }\href {\doibase 10.1038/nature21426}
  {\bibfield  {journal} {\bibinfo  {journal} {Nature}\ }\textbf {\bibinfo
  {volume} {543}},\ \bibinfo {pages} {221} (\bibinfo {year} {2017})},\ \Eprint
  {http://arxiv.org/abs/1610.08057} {arXiv:1610.08057 [quant-ph]} \BibitemShut
  {NoStop}%
\bibitem [{\citenamefont {Lazarides}\ and\ \citenamefont
  {Moessner}(2017)}]{PhysRevB.95.195135}%
  \BibitemOpen
  \bibfield  {author} {\bibinfo {author} {\bibfnamefont {A.}~\bibnamefont
  {Lazarides}}\ and\ \bibinfo {author} {\bibfnamefont {R.}~\bibnamefont
  {Moessner}},\ }\href {\doibase 10.1103/PhysRevB.95.195135} {\bibfield
  {journal} {\bibinfo  {journal} {Phys. Rev. B}\ }\textbf {\bibinfo {volume}
  {95}},\ \bibinfo {pages} {195135} (\bibinfo {year} {2017})}\BibitemShut
  {NoStop}%
\bibitem [{\citenamefont {Ho}\ \emph {et~al.}(2017)\citenamefont {Ho},
  \citenamefont {Choi}, \citenamefont {Lukin},\ and\ \citenamefont
  {Abanin}}]{PhysRevLett.119.010602}%
  \BibitemOpen
  \bibfield  {author} {\bibinfo {author} {\bibfnamefont {W.~W.}\ \bibnamefont
  {Ho}}, \bibinfo {author} {\bibfnamefont {S.}~\bibnamefont {Choi}}, \bibinfo
  {author} {\bibfnamefont {M.~D.}\ \bibnamefont {Lukin}}, \ and\ \bibinfo
  {author} {\bibfnamefont {D.~A.}\ \bibnamefont {Abanin}},\ }\href {\doibase
  10.1103/PhysRevLett.119.010602} {\bibfield  {journal} {\bibinfo  {journal}
  {Phys. Rev. Lett.}\ }\textbf {\bibinfo {volume} {119}},\ \bibinfo {pages}
  {010602} (\bibinfo {year} {2017})}\BibitemShut {NoStop}%
\bibitem [{\citenamefont {{Russomanno}}\ \emph {et~al.}(2017)\citenamefont
  {{Russomanno}}, \citenamefont {{Iemini}}, \citenamefont {{Dalmonte}},\ and\
  \citenamefont {{Fazio}}}]{2017PhRvB..95u4307R}%
  \BibitemOpen
  \bibfield  {author} {\bibinfo {author} {\bibfnamefont {A.}~\bibnamefont
  {{Russomanno}}}, \bibinfo {author} {\bibfnamefont {F.}~\bibnamefont
  {{Iemini}}}, \bibinfo {author} {\bibfnamefont {M.}~\bibnamefont
  {{Dalmonte}}}, \ and\ \bibinfo {author} {\bibfnamefont {R.}~\bibnamefont
  {{Fazio}}},\ }\href {\doibase 10.1103/PhysRevB.95.214307} {\bibfield
  {journal} {\bibinfo  {journal} {Phys. Rev. B}\ }\textbf {\bibinfo {volume}
  {95}},\ \bibinfo {eid} {214307} (\bibinfo {year} {2017})},\ \Eprint
  {http://arxiv.org/abs/1704.01591} {arXiv:1704.01591 [cond-mat.stat-mech]}
  \BibitemShut {NoStop}%
\bibitem [{\citenamefont {{Gong}}\ \emph {et~al.}(2017)\citenamefont {{Gong}},
  \citenamefont {{Hamazaki}},\ and\ \citenamefont
  {{Ueda}}}]{2017arXiv170801472G}%
  \BibitemOpen
  \bibfield  {author} {\bibinfo {author} {\bibfnamefont {Z.}~\bibnamefont
  {{Gong}}}, \bibinfo {author} {\bibfnamefont {R.}~\bibnamefont {{Hamazaki}}},
  \ and\ \bibinfo {author} {\bibfnamefont {M.}~\bibnamefont {{Ueda}}},\
  }\href@noop {} {\bibfield  {journal} {\bibinfo  {journal} {ArXiv e-prints}\ }
  (\bibinfo {year} {2017})},\ \Eprint {http://arxiv.org/abs/1708.01472}
  {arXiv:1708.01472 [cond-mat.stat-mech]} \BibitemShut {NoStop}%
\bibitem [{\citenamefont {{Flicker}}(2017{\natexlab{a}})}]{timequasicrystal}%
  \BibitemOpen
  \bibfield  {author} {\bibinfo {author} {\bibfnamefont {F.}~\bibnamefont
  {{Flicker}}},\ }\href@noop {} {\bibfield  {journal} {\bibinfo  {journal}
  {ArXiv e-prints}\ } (\bibinfo {year} {2017}{\natexlab{a}})},\ \Eprint
  {http://arxiv.org/abs/1707.09371} {arXiv:1707.09371 [nlin.CD]} \BibitemShut
  {NoStop}%
\bibitem [{\citenamefont {{Flicker}}(2017{\natexlab{b}})}]{timequasicrystal1}%
  \BibitemOpen
  \bibfield  {author} {\bibinfo {author} {\bibfnamefont {F.}~\bibnamefont
  {{Flicker}}},\ }\href@noop {} {\bibfield  {journal} {\bibinfo  {journal}
  {ArXiv e-prints}\ } (\bibinfo {year} {2017}{\natexlab{b}})},\ \Eprint
  {http://arxiv.org/abs/1707.09333} {arXiv:1707.09333 [nlin.CD]} \BibitemShut
  {NoStop}%
\bibitem [{\citenamefont {{Syrwid}}\ \emph {et~al.}(2017)\citenamefont
  {{Syrwid}}, \citenamefont {{Zakrzewski}},\ and\ \citenamefont
  {{Sacha}}}]{Syrwid2017}%
  \BibitemOpen
  \bibfield  {author} {\bibinfo {author} {\bibfnamefont {A.}~\bibnamefont
  {{Syrwid}}}, \bibinfo {author} {\bibfnamefont {J.}~\bibnamefont
  {{Zakrzewski}}}, \ and\ \bibinfo {author} {\bibfnamefont {K.}~\bibnamefont
  {{Sacha}}},\ }\href@noop {} {\bibfield  {journal} {\bibinfo  {journal} {ArXiv
  e-prints}\ } (\bibinfo {year} {2017})},\ \Eprint
  {http://arxiv.org/abs/1702.05006} {arXiv:1702.05006 [quant-ph]} \BibitemShut
  {NoStop}%
\bibitem [{\citenamefont {{Shapere}}\ and\ \citenamefont
  {{Wilczek}}(2017)}]{Shapere2017}%
  \BibitemOpen
  \bibfield  {author} {\bibinfo {author} {\bibfnamefont {A.~D.}\ \bibnamefont
  {{Shapere}}}\ and\ \bibinfo {author} {\bibfnamefont {F.}~\bibnamefont
  {{Wilczek}}},\ }\href@noop {} {\bibfield  {journal} {\bibinfo  {journal}
  {ArXiv e-prints}\ } (\bibinfo {year} {2017})},\ \Eprint
  {http://arxiv.org/abs/1708.03348} {arXiv:1708.03348 [cond-mat.stat-mech]}
  \BibitemShut {NoStop}%
\bibitem [{\citenamefont {Lipkin}\ \emph {et~al.}(1965)\citenamefont {Lipkin},
  \citenamefont {Meshkov},\ and\ \citenamefont {Glick}}]{LIPKIN1965188}%
  \BibitemOpen
  \bibfield  {author} {\bibinfo {author} {\bibfnamefont {H.}~\bibnamefont
  {Lipkin}}, \bibinfo {author} {\bibfnamefont {N.}~\bibnamefont {Meshkov}}, \
  and\ \bibinfo {author} {\bibfnamefont {A.}~\bibnamefont {Glick}},\ }\href
  {\doibase http://dx.doi.org/10.1016/0029-5582(65)90862-X} {\bibfield
  {journal} {\bibinfo  {journal} {Nuclear Physics}\ }\textbf {\bibinfo {volume}
  {62}},\ \bibinfo {pages} {188 } (\bibinfo {year} {1965})}\BibitemShut
  {NoStop}%
\bibitem [{\citenamefont {{Blin}}\ \emph {et~al.}(1996)\citenamefont {{Blin}},
  \citenamefont {{Hiller}},\ and\ \citenamefont
  {{Junqing}}}]{0305-4470-29-14-023}%
  \BibitemOpen
  \bibfield  {author} {\bibinfo {author} {\bibfnamefont {A.~H.}\ \bibnamefont
  {{Blin}}}, \bibinfo {author} {\bibfnamefont {B.}~\bibnamefont {{Hiller}}}, \
  and\ \bibinfo {author} {\bibfnamefont {L.}~\bibnamefont {{Junqing}}},\ }\href
  {\doibase 10.1088/0305-4470/29/14/023} {\bibfield  {journal} {\bibinfo
  {journal} {Journal of Physics A Mathematical General}\ }\textbf {\bibinfo
  {volume} {29}},\ \bibinfo {pages} {3993} (\bibinfo {year}
  {1996})}\BibitemShut {NoStop}%
\bibitem [{\citenamefont {Dusuel}\ and\ \citenamefont
  {Vidal}(2005)}]{Dusuel2005}%
  \BibitemOpen
  \bibfield  {author} {\bibinfo {author} {\bibfnamefont {S.}~\bibnamefont
  {Dusuel}}\ and\ \bibinfo {author} {\bibfnamefont {J.}~\bibnamefont {Vidal}},\
  }\href {\doibase 10.1103/PhysRevB.71.224420} {\bibfield  {journal} {\bibinfo
  {journal} {Phys. Rev. B}\ }\textbf {\bibinfo {volume} {71}},\ \bibinfo
  {pages} {224420} (\bibinfo {year} {2005})}\BibitemShut {NoStop}%
\bibitem [{\citenamefont {Ribeiro}\ \emph {et~al.}(2008)\citenamefont
  {Ribeiro}, \citenamefont {Vidal},\ and\ \citenamefont
  {Mosseri}}]{PhysRevE.78.021106}%
  \BibitemOpen
  \bibfield  {author} {\bibinfo {author} {\bibfnamefont {P.}~\bibnamefont
  {Ribeiro}}, \bibinfo {author} {\bibfnamefont {J.}~\bibnamefont {Vidal}}, \
  and\ \bibinfo {author} {\bibfnamefont {R.}~\bibnamefont {Mosseri}},\ }\href
  {\doibase 10.1103/PhysRevE.78.021106} {\bibfield  {journal} {\bibinfo
  {journal} {Phys. Rev. E}\ }\textbf {\bibinfo {volume} {78}},\ \bibinfo
  {pages} {021106} (\bibinfo {year} {2008})}\BibitemShut {NoStop}%
\bibitem [{\citenamefont {Dusuel}\ and\ \citenamefont
  {Vidal}(2004)}]{PhysRevLett.93.237204}%
  \BibitemOpen
  \bibfield  {author} {\bibinfo {author} {\bibfnamefont {S.}~\bibnamefont
  {Dusuel}}\ and\ \bibinfo {author} {\bibfnamefont {J.}~\bibnamefont {Vidal}},\
  }\href {\doibase 10.1103/PhysRevLett.93.237204} {\bibfield  {journal}
  {\bibinfo  {journal} {Phys. Rev. Lett.}\ }\textbf {\bibinfo {volume} {93}},\
  \bibinfo {pages} {237204} (\bibinfo {year} {2004})}\BibitemShut {NoStop}%
\bibitem [{\citenamefont {Quan}\ \emph {et~al.}(2007)\citenamefont {Quan},
  \citenamefont {Wang},\ and\ \citenamefont {Sun}}]{Quan2007}%
  \BibitemOpen
  \bibfield  {author} {\bibinfo {author} {\bibfnamefont {H.~T.}\ \bibnamefont
  {Quan}}, \bibinfo {author} {\bibfnamefont {Z.~D.}\ \bibnamefont {Wang}}, \
  and\ \bibinfo {author} {\bibfnamefont {C.~P.}\ \bibnamefont {Sun}},\ }\href
  {\doibase 10.1103/PhysRevA.76.012104} {\bibfield  {journal} {\bibinfo
  {journal} {Phys. Rev. A}\ }\textbf {\bibinfo {volume} {76}},\ \bibinfo
  {pages} {012104} (\bibinfo {year} {2007})}\BibitemShut {NoStop}%
\bibitem [{\citenamefont {Morrison}\ and\ \citenamefont
  {Parkins}(2008)}]{PhysRevLett.100.040403}%
  \BibitemOpen
  \bibfield  {author} {\bibinfo {author} {\bibfnamefont {S.}~\bibnamefont
  {Morrison}}\ and\ \bibinfo {author} {\bibfnamefont {A.~S.}\ \bibnamefont
  {Parkins}},\ }\href {\doibase 10.1103/PhysRevLett.100.040403} {\bibfield
  {journal} {\bibinfo  {journal} {Phys. Rev. Lett.}\ }\textbf {\bibinfo
  {volume} {100}},\ \bibinfo {pages} {040403} (\bibinfo {year}
  {2008})}\BibitemShut {NoStop}%
\bibitem [{\citenamefont {Ma}\ \emph {et~al.}(2017{\natexlab{a}})\citenamefont
  {Ma}, \citenamefont {Hoang}, \citenamefont {Gong}, \citenamefont {Li},\ and\
  \citenamefont {Yin}}]{MaYue2017}%
  \BibitemOpen
  \bibfield  {author} {\bibinfo {author} {\bibfnamefont {Y.}~\bibnamefont
  {Ma}}, \bibinfo {author} {\bibfnamefont {T.~M.}\ \bibnamefont {Hoang}},
  \bibinfo {author} {\bibfnamefont {M.}~\bibnamefont {Gong}}, \bibinfo {author}
  {\bibfnamefont {T.}~\bibnamefont {Li}}, \ and\ \bibinfo {author}
  {\bibfnamefont {Z.-q.}\ \bibnamefont {Yin}},\ }\href {\doibase
  10.1103/PhysRevA.96.023827} {\bibfield  {journal} {\bibinfo  {journal} {Phys.
  Rev. A}\ }\textbf {\bibinfo {volume} {96}},\ \bibinfo {pages} {023827}
  (\bibinfo {year} {2017}{\natexlab{a}})}\BibitemShut {NoStop}%
\bibitem [{\citenamefont {Zibold}\ \emph {et~al.}(2010)\citenamefont {Zibold},
  \citenamefont {Nicklas}, \citenamefont {Gross},\ and\ \citenamefont
  {Oberthaler}}]{PhysRevLett.105.204101}%
  \BibitemOpen
  \bibfield  {author} {\bibinfo {author} {\bibfnamefont {T.}~\bibnamefont
  {Zibold}}, \bibinfo {author} {\bibfnamefont {E.}~\bibnamefont {Nicklas}},
  \bibinfo {author} {\bibfnamefont {C.}~\bibnamefont {Gross}}, \ and\ \bibinfo
  {author} {\bibfnamefont {M.~K.}\ \bibnamefont {Oberthaler}},\ }\href
  {\doibase 10.1103/PhysRevLett.105.204101} {\bibfield  {journal} {\bibinfo
  {journal} {Phys. Rev. Lett.}\ }\textbf {\bibinfo {volume} {105}},\ \bibinfo
  {pages} {204101} (\bibinfo {year} {2010})}\BibitemShut {NoStop}%
\bibitem [{\citenamefont {Ma}\ \emph {et~al.}(2011)\citenamefont {Ma},
  \citenamefont {Wang}, \citenamefont {Sun},\ and\ \citenamefont
  {Nori}}]{MA201189}%
  \BibitemOpen
  \bibfield  {author} {\bibinfo {author} {\bibfnamefont {J.}~\bibnamefont
  {Ma}}, \bibinfo {author} {\bibfnamefont {X.}~\bibnamefont {Wang}}, \bibinfo
  {author} {\bibfnamefont {C.}~\bibnamefont {Sun}}, \ and\ \bibinfo {author}
  {\bibfnamefont {F.}~\bibnamefont {Nori}},\ }\href {\doibase
  http://dx.doi.org/10.1016/j.physrep.2011.08.003} {\bibfield  {journal}
  {\bibinfo  {journal} {Physics Reports}\ }\textbf {\bibinfo {volume} {509}},\
  \bibinfo {pages} {89 } (\bibinfo {year} {2011})}\BibitemShut {NoStop}%
\bibitem [{\citenamefont {Zhang}\ \emph {et~al.}(2017)\citenamefont {Zhang},
  \citenamefont {Zhou}, \citenamefont {Zhou}, \citenamefont {Guo},\ and\
  \citenamefont {Zhou}}]{PhysRevLett.118.083604}%
  \BibitemOpen
  \bibfield  {author} {\bibinfo {author} {\bibfnamefont {Y.-C.}\ \bibnamefont
  {Zhang}}, \bibinfo {author} {\bibfnamefont {X.-F.}\ \bibnamefont {Zhou}},
  \bibinfo {author} {\bibfnamefont {X.}~\bibnamefont {Zhou}}, \bibinfo {author}
  {\bibfnamefont {G.-C.}\ \bibnamefont {Guo}}, \ and\ \bibinfo {author}
  {\bibfnamefont {Z.-W.}\ \bibnamefont {Zhou}},\ }\href {\doibase
  10.1103/PhysRevLett.118.083604} {\bibfield  {journal} {\bibinfo  {journal}
  {Phys. Rev. Lett.}\ }\textbf {\bibinfo {volume} {118}},\ \bibinfo {pages}
  {083604} (\bibinfo {year} {2017})}\BibitemShut {NoStop}%
\bibitem [{\citenamefont {Or\'us}\ \emph {et~al.}(2008)\citenamefont {Or\'us},
  \citenamefont {Dusuel},\ and\ \citenamefont
  {Vidal}}]{PhysRevLett.101.025701}%
  \BibitemOpen
  \bibfield  {author} {\bibinfo {author} {\bibfnamefont {R.}~\bibnamefont
  {Or\'us}}, \bibinfo {author} {\bibfnamefont {S.}~\bibnamefont {Dusuel}}, \
  and\ \bibinfo {author} {\bibfnamefont {J.}~\bibnamefont {Vidal}},\ }\href
  {\doibase 10.1103/PhysRevLett.101.025701} {\bibfield  {journal} {\bibinfo
  {journal} {Phys. Rev. Lett.}\ }\textbf {\bibinfo {volume} {101}},\ \bibinfo
  {pages} {025701} (\bibinfo {year} {2008})}\BibitemShut {NoStop}%
\bibitem [{\citenamefont {{Zhang}}\ and\ \citenamefont
  {{Li}}(2013)}]{2013PhLA..377.1053Z}%
  \BibitemOpen
  \bibfield  {author} {\bibinfo {author} {\bibfnamefont {X.-x.}\ \bibnamefont
  {{Zhang}}}\ and\ \bibinfo {author} {\bibfnamefont {F.-l.}\ \bibnamefont
  {{Li}}},\ }\href {\doibase 10.1016/j.physleta.2013.02.040} {\bibfield
  {journal} {\bibinfo  {journal} {Physics Letters A}\ }\textbf {\bibinfo
  {volume} {377}},\ \bibinfo {pages} {1053} (\bibinfo {year}
  {2013})}\BibitemShut {NoStop}%
\bibitem [{\citenamefont {Ma}\ \emph {et~al.}(2017{\natexlab{b}})\citenamefont
  {Ma}, \citenamefont {Su},\ and\ \citenamefont {Sun}}]{PhysRevE.96.022143}%
  \BibitemOpen
  \bibfield  {author} {\bibinfo {author} {\bibfnamefont {Y.-H.}\ \bibnamefont
  {Ma}}, \bibinfo {author} {\bibfnamefont {S.-H.}\ \bibnamefont {Su}}, \ and\
  \bibinfo {author} {\bibfnamefont {C.-P.}\ \bibnamefont {Sun}},\ }\href
  {\doibase 10.1103/PhysRevE.96.022143} {\bibfield  {journal} {\bibinfo
  {journal} {Phys. Rev. E}\ }\textbf {\bibinfo {volume} {96}},\ \bibinfo
  {pages} {022143} (\bibinfo {year} {2017}{\natexlab{b}})}\BibitemShut
  {NoStop}%
\bibitem [{\citenamefont {{Sacha}}\ and\ \citenamefont
  {{Zakrzewski}}(2017)}]{Sacha2017}%
  \BibitemOpen
  \bibfield  {author} {\bibinfo {author} {\bibfnamefont {K.}~\bibnamefont
  {{Sacha}}}\ and\ \bibinfo {author} {\bibfnamefont {J.}~\bibnamefont
  {{Zakrzewski}}},\ }\href@noop {} {\bibfield  {journal} {\bibinfo  {journal}
  {ArXiv e-prints}\ } (\bibinfo {year} {2017})},\ \Eprint
  {http://arxiv.org/abs/1704.03735} {arXiv:1704.03735 [quant-ph]} \BibitemShut
  {NoStop}%
\bibitem [{\citenamefont {Bogoliubov}(1970)}]{Bogoliubov}%
  \BibitemOpen
  \bibfield  {author} {\bibinfo {author} {\bibfnamefont {N.}~\bibnamefont
  {Bogoliubov}},\ }\href@noop {} {\emph {\bibinfo {title} {Lectures on Quantum
  Statistics}}},\ Vol.~\bibinfo {volume} {2}\ (\bibinfo  {publisher} {Gordon
  and Breach},\ \bibinfo {address} {new York},\ \bibinfo {year}
  {1970})\BibitemShut {NoStop}%
\bibitem [{\citenamefont {Newman}\ and\ \citenamefont
  {Schulman}(1977)}]{Newman1977}%
  \BibitemOpen
  \bibfield  {author} {\bibinfo {author} {\bibfnamefont {C.~M.}\ \bibnamefont
  {Newman}}\ and\ \bibinfo {author} {\bibfnamefont {L.~S.}\ \bibnamefont
  {Schulman}},\ }\href {\doibase 10.1063/1.523131} {\bibfield  {journal}
  {\bibinfo  {journal} {Journal of Mathematical Physics}\ }\textbf {\bibinfo
  {volume} {18}},\ \bibinfo {pages} {23} (\bibinfo {year} {1977})}\BibitemShut
  {NoStop}%
\bibitem [{\citenamefont {Botet}\ and\ \citenamefont
  {Jullien}(1983)}]{Botet1983}%
  \BibitemOpen
  \bibfield  {author} {\bibinfo {author} {\bibfnamefont {R.}~\bibnamefont
  {Botet}}\ and\ \bibinfo {author} {\bibfnamefont {R.}~\bibnamefont
  {Jullien}},\ }\href {\doibase 10.1103/PhysRevB.28.3955} {\bibfield  {journal}
  {\bibinfo  {journal} {Phys. Rev. B}\ }\textbf {\bibinfo {volume} {28}},\
  \bibinfo {pages} {3955} (\bibinfo {year} {1983})}\BibitemShut {NoStop}%
\bibitem [{\citenamefont {{Boyle}}\ and\ \citenamefont
  {{Steinhardt}}(2016)}]{2016arXiv160808220B}%
  \BibitemOpen
  \bibfield  {author} {\bibinfo {author} {\bibfnamefont {L.}~\bibnamefont
  {{Boyle}}}\ and\ \bibinfo {author} {\bibfnamefont {P.~J.}\ \bibnamefont
  {{Steinhardt}}},\ }\href@noop {} {\bibfield  {journal} {\bibinfo  {journal}
  {ArXiv e-prints}\ } (\bibinfo {year} {2016})},\ \Eprint
  {http://arxiv.org/abs/1608.08220} {arXiv:1608.08220 [math-ph]} \BibitemShut
  {NoStop}%
\bibitem [{\citenamefont {{Giergiel}}\ \emph {et~al.}(2017)\citenamefont
  {{Giergiel}}, \citenamefont {{Miroszewski}},\ and\ \citenamefont
  {{Sacha}}}]{2017arXiv171010087G}%
  \BibitemOpen
  \bibfield  {author} {\bibinfo {author} {\bibfnamefont {K.}~\bibnamefont
  {{Giergiel}}}, \bibinfo {author} {\bibfnamefont {A.}~\bibnamefont
  {{Miroszewski}}}, \ and\ \bibinfo {author} {\bibfnamefont {K.}~\bibnamefont
  {{Sacha}}},\ }\href@noop {} {\bibfield  {journal} {\bibinfo  {journal} {ArXiv
  e-prints}\ } (\bibinfo {year} {2017})},\ \Eprint
  {http://arxiv.org/abs/1710.10087} {arXiv:1710.10087 [cond-mat.quant-gas]}
  \BibitemShut {NoStop}%
\bibitem [{\citenamefont {Song}\ \emph {et~al.}(2017)\citenamefont {Song},
  \citenamefont {Xu}, \citenamefont {Liu}, \citenamefont {Yang}, \citenamefont
  {Zheng}, \citenamefont {Deng}, \citenamefont {Xie}, \citenamefont {Huang},
  \citenamefont {Guo}, \citenamefont {Zhang}, \citenamefont {Zhang},
  \citenamefont {Xu}, \citenamefont {Zheng}, \citenamefont {Zhu}, \citenamefont
  {Wang}, \citenamefont {Chen}, \citenamefont {Lu}, \citenamefont {Han},\ and\
  \citenamefont {Pan}}]{2017arXiv170310302S}%
  \BibitemOpen
  \bibfield  {author} {\bibinfo {author} {\bibfnamefont {C.}~\bibnamefont
  {Song}}, \bibinfo {author} {\bibfnamefont {K.}~\bibnamefont {Xu}}, \bibinfo
  {author} {\bibfnamefont {W.}~\bibnamefont {Liu}}, \bibinfo {author}
  {\bibfnamefont {C.-p.}\ \bibnamefont {Yang}}, \bibinfo {author}
  {\bibfnamefont {S.-B.}\ \bibnamefont {Zheng}}, \bibinfo {author}
  {\bibfnamefont {H.}~\bibnamefont {Deng}}, \bibinfo {author} {\bibfnamefont
  {Q.}~\bibnamefont {Xie}}, \bibinfo {author} {\bibfnamefont {K.}~\bibnamefont
  {Huang}}, \bibinfo {author} {\bibfnamefont {Q.}~\bibnamefont {Guo}}, \bibinfo
  {author} {\bibfnamefont {L.}~\bibnamefont {Zhang}}, \bibinfo {author}
  {\bibfnamefont {P.}~\bibnamefont {Zhang}}, \bibinfo {author} {\bibfnamefont
  {D.}~\bibnamefont {Xu}}, \bibinfo {author} {\bibfnamefont {D.}~\bibnamefont
  {Zheng}}, \bibinfo {author} {\bibfnamefont {X.}~\bibnamefont {Zhu}}, \bibinfo
  {author} {\bibfnamefont {H.}~\bibnamefont {Wang}}, \bibinfo {author}
  {\bibfnamefont {Y.-A.}\ \bibnamefont {Chen}}, \bibinfo {author}
  {\bibfnamefont {C.-Y.}\ \bibnamefont {Lu}}, \bibinfo {author} {\bibfnamefont
  {S.}~\bibnamefont {Han}}, \ and\ \bibinfo {author} {\bibfnamefont {J.-W.}\
  \bibnamefont {Pan}},\ }\href {\doibase 10.1103/PhysRevLett.119.180511}
  {\bibfield  {journal} {\bibinfo  {journal} {Phys. Rev. Lett.}\ }\textbf
  {\bibinfo {volume} {119}},\ \bibinfo {pages} {180511} (\bibinfo {year}
  {2017})}\BibitemShut {NoStop}%
\end{thebibliography}

%

\end{document}